\documentclass[%
 reprint,
 amsmath,amssymb,
 aps,
]{revtex4-2}
\newcommand{\PiT}{\Pi_{\text{T}}}
\newcommand{\PiL}{\Pi_{\text{L}}}
\newcommand{\PT}{P_{\text{T}}}
\newcommand{\PL}{P_{\text{L}}}

\usepackage{graphicx}% Include figure files
\usepackage{dcolumn}% Align Chabowski:2024smx columns on decimal point
\usepackage{bm}% bold math
\usepackage{color}
\usepackage[normalem]{ulem}
\begin{document}

\preprint{APS/123-QED}

\title{Polarization dependence of the $\phi$ meson from finite-temperature QCD sum rules
}% Force line breaks with \\

\author{Hidefumi Matsuda}
\email{da.matsu.00.bbb.kobe@gmail.com}
\affiliation{Zhejiang Institute of Modern Physics, Department of Physics, Zhejiang University, Hangzhou, 310027, China}

\author{Philipp Gubler}
\email{gubler.philipp@jaea.go.jp; philipp.gubler1@gmail.com}
\affiliation{Advanced Science Research Center, Japan Atomic Energy Agency, Tokai, Ibaraki 319-1195, Japan}

\author{Koichi Hattori}
\email{koichi.hattori@zju.edu.cn}
\affiliation{Zhejiang Institute of Modern Physics, Department of Physics, Zhejiang University, Hangzhou, 310027, China}
\affiliation{Research Center for Nuclear Physics (RCNP), Osaka University, Osaka 567-0047, Japan}


\begin{abstract}
We study the $\phi$ meson at finite temperature and finite momentum using QCD sum rules. 
The presence of medium breaks the Lorentz invariance, and induces distinct in-medium modifications of the transverse and longitudinal modes at finite momentum.  
We find that, with increasing momentum, the masses of both modes increase 
and a clear transverse--longitudinal splitting develops.
The splitting is found to grow with temperature and to be mainly generated by the
dimension-four spin-dependent thermal condensates.
\end{abstract}

\maketitle

%\Chabowski:2024smxofcontents

\section{Introduction}
In-medium effects on vector mesons serve as a useful experimental probe of the low-temperature and low-to-intermediate-density region 
of the QCD phase diagram. 
On the timescale of hadronic matter created in collision experiments, 
vector mesons are better regarded as finite-lifetime impurities propagating through the medium rather than as particles fully equilibrated with it. 
Their properties are expected to be modified by interactions with the surrounding hadronic environment.
Such in-medium modifications of vector meson properties can reflect essential features of the hadronic regime of the QCD phase diagram, 
such as the partial restoration of chiral symmetry, medium-induced hadronic excitations, 
and temperature- and density-dependent changes in the dominant hadronic degrees of freedom of the medium.

The $\phi$ meson occupies a special position among light vector mesons since its quark 
content is close to a pure $s\bar{s}$ state~\cite{Dudek:2013yja}, and its in-medium properties are strongly 
influenced by strange hadronic channels, especially the $K\bar{K}$ channel. 
It therefore provides information complementary to that obtained from the $\rho$ and 
$\omega$ mesons. 
Experimental measurements and related analyses at various facilities have provided evidence for in-medium modifications of the $\phi$meson properties, in particular those induced by finite-density effects~\cite{Ishikawa:2004id,KEK-PS-E325:2005wbm,E325:2006ioe,CLAS:2010pxs,Polyanskiy:2010tj,Hartmann:2012ia,HADES:2018qkj,KEK-PSE325:2025fms}. While these studies have mainly addressed finite-density effects, the focus has more recently extended to regimes where finite-temperature effects also become important. Ongoing and planned experimental programs, such as J-PARC-HI, CBM at FAIR, MPD at NICA, and CEE at HIRFL-CSR, aim to explore such regimes of the QCD phase diagram.

Theoretical studies of $\phi$ meson properties at finite temperature have a long history. 
Important results have been obtained from QCD sum rules~\cite{Asakawa:1994tp,Zschocke:2002mn} 
as well as from hadronic effective-model approaches 
which include effects of fluctuations in relevant hadronic channels, such as $K\bar{K}$ and $\pi\rho$, and scattering processes with thermal hadrons in the medium~\cite{Lissauer:1991fr,Ko:1993id,Haglin:1994ap,Song:1996gw,Vujanovic:2009wr,Kumar:2020vys,Kaur:2025kjk}.
It is challenging, in particular, for lattice QCD simulations because extracting the real-time spectral information at finite temperature requires analytic continuation from the imaginary-time formalism, even at zero chemical potential where the fermion sign problem does not arise.

The meson properties depend not only on temperature but also on momentum because of the existence of a preferred Lorentz frame, i.e., a medium rest frame. 
Understanding the momentum dependence is important for phenomenological applications to the aforementioned experiments.
Furthermore, the absence of Lorentz invariance implies that 
the longitudinal and transverse modes with respect to a vector-meson momentum can receive different modifications at finite momentum. 
As finite-temperature regimes become experimentally accessible and precision in experimental measurements is improved, 
this effect is becoming increasingly relevant~\cite{Park:2022ayr,Arifi:2026eus}. 
Investigating the thermal origin of the longitudinal-transverse splitting of the finite-momentum $\phi$ meson is therefore a timely theoretical task, also in light of the recent measurements of the $\phi$ meson polarization in non-central heavy-ion collisions ~\cite{ALICE:2019aid,STAR:2022fan}, for which the polarization dependence considered in this work could be relevant.

In this paper, we study the in-medium properties of the $\phi$ meson
in the low-temperature hadronic phase at vanishing baryon number using QCD sum rules~\cite{Shifman:1978bx,Shifman:1978by} as a nonperturbative method complementary to lattice QCD simulations.
Our main focus is the longitudinal-transverse splitting of the $\phi$ meson 
at finite momentum, and we systematically investigate its temperature and momentum dependence. 
The QCD sum rules has been successfully applied to the $\phi$ meson 
in vacuum~\cite{Shifman:1978by} and in medium~\cite{Hatsuda:1991ez,Asakawa:1994tp,Zschocke:2002mn,Klingl:1997kf,Lee:1997zta,Kim:2019ybi}. 
We include the non-scalar condensates up to dimension six in the OPE following Ref.~\cite{Kim:2017nyg}, capturing both the leading ($d=4$) and subleading ($d=6$) contributions to the splitting.
We evaluate the non-scalar condensates, which are purely medium-induced contributions, using the dilute pion gas approximation~\cite{Hatsuda:1992bv}.
This is justified because pions are the relevant degrees of freedom in our target temperature range. 
With these ingredients, we identify the non-scalar condensates responsible for it.

This paper is organized as follows.
In Sec.~\ref{sec:qcd_sum_rules}, we introduce the basic framework of QCD sum rules 
and the explicit setup used in the present analysis.
In Sec.~\ref{sec:results}, we present numerical results for the $\phi$ meson 
properties at finite temperature.
In Sec.~\ref{sec:summary}, we summarize our findings.

\section{QCD sum rules}
\label{sec:qcd_sum_rules}

In this section, we first recaptulate basics of QCD sum rules applied to finite temperature and the OPE developed in Ref.~\cite{Kim:2017nyg}. 
Then, we evaluate the relevant condensates at finite temperature in the subsections after Sec.~\ref{subsec:input_scalar_condensates}. 
We also summarize the phenomenological modeling of the spectral function and the Borel transformation for the numerical analysis performed in the result section.

\subsection{Basic idea of the QCD sum rule}
\label{subsec:qcdsr_basic}
QCD sum rules are a QCD-based approach to extracting hadronic spectral information from two-point correlation functions. This approach relies on a dispersion relation connecting the correlation function to an integral of the spectral function over the invariant mass. The correlator is evaluated at deep spacelike momenta using the OPE, while the spectral function 
is parametrized phenomenologically (see, however, Refs.\cite{Gubler:2010cf,Li:2020ejs} for attempts without explicit parametrizations).
In this sense, the dispersion relation serves as a bridge between the 
QCD side, where the short-distance physics is encoded in the OPE, and 
the hadronic side, where the long-distance physics is represented by the 
spectral function. The model parameters, such as the hadron mass and 
hadron decay width, are then determined by requiring consistency between the two sides 
of the dispersion relation.

The present analysis for the $\phi$ meson is based on the thermal current
correlator 
%of the strange vector current,
\begin{align}
  \Pi^{\mu\nu}(q,T)
  =
  i\int d^4x\, e^{iq\cdot x}
  \left\langle
    \mathcal{T} J^\mu(x)J^\nu(0)
  \right\rangle_T\ ,
  \label{eq:thermal_correlator}
\end{align}
where \(J^\mu=\bar{s}\gamma^\mu s\) is the strange vector current and
\(\langle\cdots\rangle_T\) denotes the thermal average.
The correlator $\Pi^{\mu\nu}$ is symmetric and, by the Ward identity, 
transverse to the four-momentum: $q_\mu \Pi^{\mu\nu} = 0$. 
In vacuum, the only available symmetric second-rank tensors are $q^\mu q^\nu$ 
and $g^{\mu\nu}$, and these conditions reduce $\Pi^{\mu\nu}$ to a single 
scalar amplitude, 
$\Pi^{\mu\nu} = P^{\mu\nu} \Pi^{\mathrm{vac}}$ where $P^{\mu\nu}  = q^\mu q^\nu - q^2 g^{\mu\nu}$. 
At finite temperature, there is a timelike vector $u^\mu$, of which the spatial component specifies the collective velocity of medium. 
In particular, the medium rest frame is specified as $u^\mu_{\rm rest} = (1,\mathbf{0})$. 
The existence of this external vector breaks the Lorentz symmetry, and $\Pi^{\mu\nu}$ now admits two independent tensor structures transverse and longitudinal to $u^\mu $, which will be shown in Eq.~(\ref{eq:TL_decomposition}).

The existence of $u^\mu$ splits $P^{\mu\nu}$ into the transverse and longitudinal components such that 
\begin{eqnarray}
u_\mu P_T^{\mu\nu} = 0 , \quad 
u_\mu P_L^{\mu\nu} \not= 0 ,
\end{eqnarray} 
where $ \PL^{\mu\nu} +  \PT^{\mu\nu} = P^{\mu\nu}$. 
Both of them remain transverse to the four-momentum and are orthogonal to each other 
\begin{align}
  q_\mu \PT^{\mu\nu}
  &=
  q_\mu \PL^{\mu\nu}
  =
  0\ ,
  \\
{\PT^\mu}_{\alpha}\PL^{\alpha\nu} 
  &= 0 \ .
\end{align}
In the medium rest frame, the transverse component reads 
\begin{align}
  \PT^{00}
  &=
  \PT^{0i}
  =
  \PT^{i0}
  =
  0\ ,
  \qquad
  \PT^{ij}
  =
  q^2\,\delta^{ij}
  -
  q^{\,i}q^{\,j}\ .
  \label{eq:transverse_projector}
\end{align}
Note that $ \PT^{ij}$ is transverse to spatial momentum $q^i$, while $ \PL^{ij}$ is not. 
Therefore, they project out polarization modes in the medium rest frame.   

By the use of the projection operators, the correlator is decomposed as
\begin{align}
  \Pi^{\mu\nu}
  =
  \PT^{\mu\nu}\PiT
  +
  \PL^{\mu\nu}\PiL\ .
  \label{eq:TL_decomposition}
\end{align}
Conversely, the scalar amplitudes are extracted as
\begin{align}
  \PiL(q^2,v^2)
  &=
  \frac{1}{v^2}
  \Pi_{00}(q^2,v^2),
  \label{eq:PiL_extraction}
  \\
  \PiT(q^2,v^2)
  &=
  -\frac{1}{2}
  \left[
    \frac{1}{q^2}{\Pi^\mu}_{\mu}(q^2,v^2)
    +
    \PiL(q^2,v^2)
  \right]\ ,
  \label{eq:PiT_extraction}
\end{align}
where the scalar amplitudes are expressed as functions of $q^2$ and $v^2 \equiv (u\cdot q)^2 - q^2$.
Since the projection operators contain inverse powers of $v$, these expressions are 
not evaluated directly at $v=0$; instead, the projection is carried out at 
finite $v$, and then the limit $v\to0$ is taken. In this limit, $q^\mu$ becomes proportional to $u^\mu$, 
and the two amplitudes reduce to one, $\PiT = \PiL$. In the following, we set $u^\mu = u^\mu_{\rm rest}$. 

For each scalar amplitude, a dispersion relation is given as
\begin{align}
  \Pi_{\text{X}}(q^2,v^2)
  =
  \int_{-v^2}^{\infty} ds\,
  \frac{\rho_{\text{X}}(s,v^2)}{s-q^2}\ ,
  \qquad
  (\text{X}=\text{L},\text{T})\ ,
  \label{eq:reduced_dispersion_relation}
\end{align}
where $\rho_{\text{X}}$ is the spectral function, related to the Feynman 
correlator by
\begin{align}
  \rho_{\text{X}}(q^2, v^2) = \frac{1}{\pi}\tanh\!\left(\frac{\beta q^0}{2}\right)
  \mathrm{Im}\,\Pi_{\text{X}}(q^2, v^2)
  \qquad (q^0 \in \mathbb{R})\ .\label{Eq:PiF_rho}
\end{align}
Here the positive energy condition $q^0>0$ uniquely gives $q^0$ as a function of $q^2$ and $v^2$.

In Eq.~\eqref{eq:reduced_dispersion_relation}, we have suppressed the terms that vanish under the Borel transformation introduced later. 
%have already been omitted. 
More explicitly, the dispersion relation is derived by applying Cauchy's 
theorem to the correlator, which yields an integral over the spectral 
function and a surface contribution from the contour at infinity. 
Both terms are in general divergent, and the divergences are removed by 
the corresponding subtraction terms.
Equation~\eqref{eq:reduced_dispersion_relation} contains only the 
subtracted integral over the spectral function. 
The subtracted surface contribution is a polynomial in $q^2$ with 
positive powers, which vanishes under the Borel transformation. 
Note that when $\Pi_{\text{X}}$ carries a thermal contribution at $q^2=0$, an additional term that survives the Borel transformation appears in the dispersion relation; we assume this contribution to be subleading, as discussed in Sec.~\ref{subsec:ch4_hadronic_spectral_ansatz}.
For details, see Ref.~\cite{Klingl:1997kf,Leupold:1998bt}.

\subsection{Operator Product Expansion up to Dimension Six}
\label{subsec:ope_dim6}

The left-hand side of Eq.~\eqref{eq:reduced_dispersion_relation} is 
evaluated using the OPE. 
In the deep Euclidean limit $Q^2\equiv -q^2\to\infty$, 
the correlator admits a systematic expansion in local operators,
\begin{equation}
\Pi(q^2,v^2)
=
\sum_n 
C_n^{\mu_1\cdots\mu_{s_n}}(q^2,v^2)\,
\left \langle \mathcal{O}^{(n)}_{\mu_1\cdots\mu_{s_n}}\right\rangle_T\ ,
\end{equation}
where the Wilson coefficients $C_n$ encode the short-distance physics and 
are calculable in perturbation theory, while the local operators 
$\mathcal{O}^{(n)}$ capture the nonperturbative long-distance effects.
The expansion proceeds on an order-by-order basis in operator dimension, with 
higher-dimensional operators suppressed by inverse powers of $Q^2$ in the corresponding Wilson coefficients.

We evaluate the OPE of the correlator up to dimension-six operators. 
The Wilson coefficients used here are derived in Ref.~\cite{Kim:2017nyg}. 
For convenience, we classify the OPE contributions by operator dimension and spin,
\begin{align}
  \Pi_{\mu\nu}^{\rm OPE}
  =
  \Pi_{\mu\nu}^{\rm scalar}
  +
  \Pi_{\mu\nu}^{4,2}
  +
  \Pi_{\mu\nu}^{6,2}
  +
  \Pi_{\mu\nu}^{6,4}\ ,
  \label{eq:ope_tensor_decomposition}
\end{align}
where $\Pi_{\mu\nu}^{\rm scalar}$ collects the scalar operator contributions 
and $\Pi_{\mu\nu}^{d,s}$ denotes the contribution of operators with 
dimension $d$ and spin $s$.
The non-scalar condensates vanish in vacuum by the Lorentz symmetry and are therefore purely thermal in origin.

Each part has the Lorentz structure
\begin{align}
\Pi^{\mathrm{scalar}}_{\mu\nu}
&=
(q_\mu q_\nu-q^2 g_{\mu\nu})\,\Pi^{\mathrm{scalar}}
 ,\\
\Pi^{4,2}_{\mu\nu}
&=
\frac{1}{Q^2}
\Bigl[
I^{4,2}_{\mu\nu}
+\frac{1}{Q^2}
\left(
q^\rho q_\mu I^{4,2}_{\rho\nu}
+
q^\rho q_\nu I^{4,2}_{\rho\mu}
\right)
\nonumber\\
&+g_{\mu\nu}\frac{q^\rho q^\sigma}{Q^2}J^{4,2}_{\rho\sigma}
\nonumber\\
&+\frac{q_\mu q_\nu q^\rho q^\sigma}{Q^4}
\left(
I^{4,2}_{\rho\sigma}
+
J^{4,2}_{\rho\sigma}
\right)
\Bigr]\ ,\\
\Pi^{6,2}_{\mu\nu}
&=
\frac{1}{Q^4}
\Bigl[
I^{6,2}_{\mu\nu}
+\frac{1}{Q^2}
\left(
q^\rho q_\mu I^{6,2}_{\rho\nu}
+
q^\rho q_\nu I^{6,2}_{\rho\mu}
\right)
\nonumber\\
&+g_{\mu\nu}\frac{q^\rho q^\sigma}{Q^2}J^{6,2}_{\rho\sigma}
\nonumber\\
&+\frac{q_\mu q_\nu q^\rho q^\sigma}{Q^4}
\left(
I^{6,2}_{\rho\sigma}
+
J^{6,2}_{\rho\sigma}
\right)
\Bigr]\ ,\\
\Pi^{6,4}_{\mu\nu}
&=
\frac{q^\kappa q^\lambda}{Q^6}
\Bigl[
I^{6,4}_{\kappa\lambda\mu\nu}
+\frac{1}{Q^2}
\left(
q^\rho q_\mu I^{6,4}_{\kappa\lambda\rho\nu}
+
q^\rho q_\nu I^{6,4}_{\kappa\lambda\rho\mu}
\right)
\nonumber\\
&+g_{\mu\nu}\frac{q^\rho q^\sigma}{Q^2}J^{6,4}_{\kappa\lambda\rho\sigma}
\nonumber\\
&+\frac{q_\mu q_\nu q^\rho q^\sigma}{Q^4}
\left(
I^{6,4}_{\kappa\lambda\rho\sigma}
+
J^{6,4}_{\kappa\lambda\rho\sigma}
\right)
\Bigr]\ ,
\end{align}
where $I^{d,s}$, $J^{d,s}$ are shorthand tensors 
defined in Ref.~\cite{Kim:2017nyg}, whose explicit expressions in terms 
of the condensates are given below.

The scalar amplitude $\Pi^{\mathrm{scalar}}$ receives contributions from 
the unit operator, which gives the purely perturbative part, and from 
scalar condensates. 
We expand the perturbative part in powers of $\alpha_s$ and the strange 
quark mass $m_s$, retaining terms up to $O(\alpha_s)$ and $O(m_s^4)$,
\begin{align}
&\Pi^{\mathrm{scalar}}_{\mathrm{pert}}
=
-\frac{1}{4\pi^2}
\left(
1+\frac{\alpha_s}{\pi}
\right)
\ln\frac{Q^2}{\mu^2}
\nonumber\\
&
-\frac{3m_s^2}{2\pi^2 Q^2}
-\frac{m_s^2}{\pi^2}
\frac{\alpha_s}{\pi}
\frac{1}{Q^2}
\left(
4
-
3\ln\frac{Q^2}{\mu^2}
\right)
\nonumber\\
&
+\frac{m_s^4}{Q^4}
\Biggl[
\frac{3}{4\pi^2}
\left(
2\ln\frac{Q^2}{\mu^2}
-
1
\right)
\nonumber\\
&-
\frac{1}{6\pi^2}
\frac{\alpha_s}{\pi}
\Bigl\{
32
-
24\zeta(3)
-
33\ln\frac{Q^2}{\mu^2}
+
18\left(\ln\frac{Q^2}{\mu^2}\right)^2
\Bigr\}
\Biggr]\ .
\label{eq:deltaPi_before_borel}
\end{align}
For the scalar-condensate contributions, we include terms up to first 
order in $\alpha_s$ and third order in $m_s$; terms of order $m_s^4$ are 
absent since they would contribute beyond dimension six,
\begin{align}
\Pi^{\mathrm{scalar}}_{\mathrm{cond}}(Q^2)
&=
\Pi^{\mathrm{scalar}}_{\langle \bar{s}s\rangle}(Q^2)
+
\Pi^{\mathrm{scalar}}_{\langle G^2\rangle}(Q^2)
+
\Pi^{\mathrm{scalar}}_{\mathrm{dim}\,6}(Q^2)\ ,
\end{align}
where
\begin{align}
\Pi^{\mathrm{scalar}}_{\langle \bar{s}s\rangle}&=
\frac{2m_s\langle \bar{s}s\rangle}{Q^4}
+
\frac{1}{Q^4}
\frac{2m_s\alpha_s}{3\pi}
\langle \bar{s}s\rangle
-
\frac{8m_s^3\langle \bar{s}s\rangle}{3Q^6}\ ,
\\
\Pi^{\mathrm{scalar}}_{\langle G^2\rangle}
&=
\frac{1}{Q^4}
\frac{7\alpha_s}{288\pi^4}
\langle G^2\rangle
+
\frac{1}{\pi^2Q^4}
\left(
\frac{1}{48}
+
\frac{1}{36}\frac{m_s^2}{Q^2}
\right)
\langle G^2\rangle\ ,
\\
\Pi^{\mathrm{scalar}}_{\mathrm{dim}\,6}
&=
-
\frac{4\langle \bar{s}js\rangle}{9Q^6}
-
\frac{2\langle j_5^2\rangle}{Q^6}
\nonumber\\
&+
\frac{1}{\pi^2Q^6}
\left(
\frac{1}{324}
+
\frac{1}{54}\ln\frac{Q^2}{\mu^2}
\right)
\langle j^2\rangle\ .
\end{align}
The scalar condensates appearing above are defined as in 
Ref.~\cite{Kim:2017nyg},
\begin{align}
\langle G^2 \rangle
&\equiv
\left\langle
g^2 G^a_{\mu\nu} G^{a\,\mu\nu}
\right\rangle\ ,
\\
\langle \bar s j s \rangle
&\equiv
\left\langle
g \bar s \gamma^\mu (D^\nu G_{\mu\nu}) s
\right\rangle\ ,
\\
\langle j_5^2 \rangle
&\equiv
\left\langle
g^2 \bar s t^a \gamma_5 \gamma^\mu s \,
\bar s t^a \gamma_5 \gamma_\mu s
\right\rangle\ ,
\\
\langle j^2 \rangle
&\equiv
\left\langle
g^2 (D^\mu G^a_{\alpha\mu})(D^\nu G^{a\,\alpha}{}_{\nu})
\right\rangle\ .
\end{align}

The Wilson coefficients of the non-scalar condensates are retained at 
leading order in $\alpha_s$. 
The quark sector contributes through
\begin{align}
I^{4,2}_{\mu\nu}
&=
\left(
4-15\frac{m_s^2}{Q^2}
\right) F_{\mu\nu}\ ,\\
J^{4,2}_{\mu\nu}
&=
\left(
-4+9\frac{m_s^2}{Q^2}
\right) F_{\mu\nu}\ ,\\
I^{6,2}_{\mu\nu}
&=
\frac{5}{2}A_{\mu\nu}
-\frac{1}{2}B_{\mu\nu}
-13 C_{\mu\nu}
+4 H_{\mu\nu}\ ,
\\
J^{6,2}_{\mu\nu}
&=
-\frac{3}{2}A_{\mu\nu}
+\frac{7}{2}B_{\mu\nu}
-5 C_{\mu\nu}
-4 H_{\mu\nu}\ ,
\\
I^{6,4}_{\mu\nu\kappa\lambda}
&=
-16 i K_{\mu\nu\kappa\lambda},
\quad
J^{6,4}_{\mu\nu\kappa\lambda}
=
16 i K_{\mu\nu\kappa\lambda}\ ,
\end{align}
with the non-scalar quark condensates defined as in Ref.~\cite{Kim:2017nyg},
\begin{align}
A_{\alpha\beta}
&\equiv
\left\langle\mathcal{ST}\,
g \bar s (D^\mu G_{\alpha\mu}) \gamma_\beta s
\right\rangle,
\\
B_{\alpha\beta}
&\equiv
\left\langle\mathcal{ST}\,
g \bar s \{ iD_\alpha , \tilde G_{\beta\mu} \}
\gamma_5 \gamma^\mu s
\right\rangle,
\\
C_{\alpha\beta}
&\equiv
\left\langle\mathcal{ST}\,
m_s \bar s D_\alpha D_\beta s
\right\rangle,
\\
F_{\alpha\beta}
&\equiv
\left\langle\mathcal{ST}\,
\bar s \gamma_\alpha iD_\beta s
\right\rangle,
\\
H_{\alpha\beta}
&\equiv
\left\langle\mathcal{ST}\,
g^2 \bar s t^a \gamma_5 \gamma_\alpha s \,
\bar s t^a \gamma_5 \gamma_\beta s
\right\rangle,
\\
K_{\alpha\beta\gamma\delta}
&\equiv
\left\langle\mathcal{ST}\,
\bar s \gamma_\alpha D_\beta D_\gamma D_\delta s
\right\rangle,
\end{align}
where $\tilde G_{\alpha\beta}=\frac{1}{2}\epsilon_{\alpha\beta\mu\nu}G^{\mu\nu}$
and $\mathcal{ST}$ denotes the symmetric traceless projection.
The gluon sector contributes through
\begin{align}
I^{4,2}_{\mu\nu}
&=
\left[
\frac{1}{8\pi^2}
-\frac{47}{48\pi^2}\frac{m_s^2}{Q^2}
+
\left(
-\frac{1}{6\pi^2}
+\frac{5}{6\pi^2}\frac{m_s^2}{Q^2}
\right)
\ln\frac{Q^2}{\mu^2}
\right]
G_{2\mu\nu}\ ,
\\
J^{4,2}_{\mu\nu}
&=
\left[
-\frac{7}{24\pi^2}
+\frac{25}{48\pi^2}\frac{m_s^2}{Q^2}
+
\left(
\frac{1}{6\pi^2}
+\frac{1}{6\pi^2}\frac{m_s^2}{Q^2}
\right)
\ln\frac{Q^2}{\mu^2}
\right]
G_{2\mu\nu}\ ,
\\
I^{6,2}_{\mu\nu}
&=
\left(
-\frac{1}{60\pi^2}
+\frac{1}{96\pi^2}\ln\frac{Q^2}{\mu^2}
\right)X_{\mu\nu}
\nonumber\\
&+
\left(
-\frac{361}{2880\pi^2}
+\frac{3}{32\pi^2}\ln\frac{Q^2}{\mu^2}
\right)Y_{\mu\nu}
\nonumber\\
&+
\left(
\frac{19}{320\pi^2}
+\frac{1}{32\pi^2}\ln\frac{Q^2}{\mu^2}
\right)Z_{\mu\nu}\ ,
\\
J^{6,2}_{\mu\nu}
&=
\left(
-\frac{1}{20\pi^2}
-\frac{7}{96\pi^2}\ln\frac{Q^2}{\mu^2}
\right)X_{\mu\nu}
\nonumber\\
&+
\left(
\frac{149}{960\pi^2}
+\frac{1}{96\pi^2}\ln\frac{Q^2}{\mu^2}
\right)Y_{\mu\nu}
\nonumber\\
&+
\left(
-\frac{239}{960\pi^2}
-\frac{7}{32\pi^2}\ln\frac{Q^2}{\mu^2}
\right)Z_{\mu\nu}\ ,
\\
I^{6,4}_{\mu\nu\kappa\lambda}
&=
\left(
-\frac{133}{180\pi^2}
+\frac{11}{30\pi^2}\ln\frac{Q^2}{\mu^2}
\right)
G_{4\mu\nu\kappa\lambda}\ ,\\
J^{6,4}_{\mu\nu\kappa\lambda}
&=
\left(
\frac{181}{180\pi^2}
-\frac{11}{30\pi^2}\ln\frac{Q^2}{\mu^2}
\right)
G_{4\mu\nu\kappa\lambda}\ ,
\end{align}
with the non-scalar gluon condensates defined as in Ref.~\cite{Kim:2017nyg},
\begin{align}
G_{2\alpha\beta}
&\equiv
\left\langle\mathcal{ST}\,
g^2 G^a_{\alpha\mu} G^{a}_{\beta}{}^\mu
\right\rangle\ ,\\
X_{\alpha\beta}
&\equiv
\left\langle\mathcal{ST}\,
g^2 G^a_{\mu\nu} D_\beta D_\alpha G^{a\,\mu\nu}
\right\rangle\ ,\\
Y_{\alpha\beta}
&\equiv
\left\langle\mathcal{ST}\,
g^2 G^a_{\alpha\mu} D^\mu D^\nu G^a_{\beta\nu}
\right\rangle,
\\
Z_{\alpha\beta}
&\equiv
\left\langle\mathcal{ST}\,
g^2 G^a_{\alpha\mu} D_\beta D^\nu G^{a\,\mu}{}_{\nu}
\right\rangle\ ,\\
G_{4\alpha\beta\gamma\delta}
&\equiv
\left\langle\mathcal{ST}\,
g^2 G^a_{\alpha\mu} D_\delta D_\gamma G^{a}_{\beta}{}^\mu
\right\rangle\ .
\end{align}
The numerical values of all input parameters and condensates are 
summarized in table~\ref{tab:ope_input_parameters}, with further details 
given in Secs.~\ref{subsec:input_scalar_condensates} 
and~\ref{subsec:input_nonscalar_condensates}.

\begin{table}[tbp]
  \centering
  \caption{
  Input parameters and condensates used in the OPE analysis.  All
  scale-dependent quantities are quoted at \(\mu=1~\mathrm{GeV}\), except for
  \(R_g^{(\pi)}\), for which we use the available lattice-QCD value as a
  reference input.
  }
  \label{tab:ope_input_parameters}
  \begin{tabular}{lll}
    \hline
    Quantity & Value & Reference \\
    \hline
    \(m_\pi\) & \(139.57061~\mathrm{MeV}\) & \cite{PDG2024} \\
    \(m_q^{\overline{\rm MS}}\) & \(4.78~\mathrm{MeV}\) & \cite{PDG2024} \\
    \(m_s^{\overline{\rm MS}}\) & \(128~\mathrm{MeV}\) & \cite{PDG2024} \\
    \(\alpha_s(1~\mathrm{GeV})\) & \(0.50\) & \cite{Bethke:2009jm} \\
    \(\langle\bar q q\rangle_0\) & \(-[232(60)~\mathrm{MeV}]^3\) & \cite{RBC:2010qam} \\
    \(\langle\bar s s\rangle_0\) & \(-[196(11)~\mathrm{MeV}]^3\) & \cite{Dominguez:2007hc} \\
    \(G_0(0)\) & \((4\pi^2)(0.012\pm0.004)~\mathrm{GeV}^4\) & \cite{Shifman:1978bx,Shifman:1978by} \\
    \(T_c^{(l)},h_l\) & \(166.762~\mathrm{MeV},\ 0.921635~\mathrm{MeV}\) & \cite{Bazavov:2011nk,Chabowski:2024smx} \\
    \(T_c^{(s)},h_s\) & \(204.316~\mathrm{MeV},\ 2.6074~\mathrm{MeV}\) & \cite{Bazavov:2011nk,Chabowski:2024smx} \\
    \(h_0,h_1,h_2\) & \(0.1396,\ -0.1800,\ 0.0350\) & \cite{Borsanyi:2013bia} \\
    \(f_0,f_1,f_2\) & \(2.76,\ 6.79,\ -5.29\) & \cite{Borsanyi:2013bia} \\
    \(g_1,g_2\) & \(-0.47,\ 1.04\) & \cite{Borsanyi:2013bia} \\
    \(A_2^{\pi(s)}\) & \(0.0257\) & \cite{Gubler:2018ctz} \\
    \(A_2^{\pi(g)}\) & \(0.380\) & \cite{Gubler:2018ctz} \\
    \(A_4^{\pi(s)}\) & \(0.00154\) & \cite{Gubler:2018ctz} \\
    \(A_4^{\pi(g)}\) & \(0.0593\) & \cite{Gubler:2018ctz} \\
    \(R_g^{(\pi)}\) & \(0.388(49)\) & \cite{ExtendedTwistedMass:2024kjf} \\
    \hline
  \end{tabular}
\end{table}

\subsection{Scalar Condensates}
\label{subsec:input_scalar_condensates}

In this subsection, we provide the numerical inputs for the scalar condensates entering the OPE, given in Sec.~\ref{subsec:ope_dim6}.
Unless otherwise stated, all
renormalized quantities are evaluated at $\mu=1~\mathrm{GeV}$ in the
$\overline{\mathrm{MS}}$ scheme.

\subsubsection{Quark Scalar Condensate: $\langle \bar{q}q\rangle$ and $\langle \bar{s}s\rangle$}
We first specify the vacuum values of quark condensates. 
We take them from lattice determinations for the light quark~\cite{RBC:2010qam} and the strange quark~\cite{Dominguez:2007hc}.
Since these values are given at $\mu = 2~\mathrm{GeV}$ in the original references, 
we convert them to $\mu = 1~\mathrm{GeV}$ using renormalization-group-invariant combinations and the running of the quark mass.
The numerical values used in the present analysis are listed in table~\ref{tab:ope_input_parameters}.
Here the light-quark condensate is evaluated in the isospin-symmetric
limit, where the up and down quarks are taken to be degenerate with a
common mass identified with the average of the physical up and down quark
masses.

The temperature dependence of the quark condensates is then captured by the smooth 
parametrizations fitted to the HotQCD data~\cite{Bazavov:2011nk,Chabowski:2024smx},
\begin{align}
  \frac{\langle \bar{q}q\rangle_T}{\langle \bar{q}q\rangle_0}
  &=
  \tanh\!\left[
    \frac{1}{T}
    \left(
      \frac{\langle \bar{q}q\rangle_T}{\langle \bar{q}q\rangle_0} \, T_c^{(l)}
      + h_l
    \right)
  \right]\ ,\label{Eq:para_1}\\
  \frac{\langle \bar{s}s\rangle_T}{\langle \bar{s}s\rangle_0}
  &=
  \tanh\!\left[
    \frac{1}{T}
    \left(
      \frac{\langle \bar{s}s\rangle_T}{\langle \bar{s}s\rangle_0} \, T_c^{(s)}
      + h_s
    \right)
  \right]\ ,\label{Eq:para_2}
\end{align}
with fit parameters
\begin{align}
  &T_c^{(l)} = 166.762~\mathrm{MeV}\ ,
  \qquad
  h_l = 0.921635~\mathrm{MeV}\ ,\\
  &T_c^{(s)} = 204.316~\mathrm{MeV}\ ,
  \qquad
  h_s = 2.6074~\mathrm{MeV}\ .
\end{align}
Since these equations are implicit, the thermal
ratios must be obtained by solving them self-consistently at each temperature.

\subsubsection{Gluon Scalar Condensate: $\langle G^2\rangle$}

For the gluon scalar condensate, we specify the vacuum value and the
thermal shift separately.
The vacuum value is taken from the standard determination, obtained within the QCD sum rule 
framework~\cite{Shifman:1978bx,Shifman:1978by},
\begin{align}
  \langle G^2\rangle_0
  =
  (4\pi^2)\,
  (0.012\pm0.004)\,\mathrm{GeV}^4\ .
  \label{eq:ope_gluon_condensate_vacuum_value}
\end{align}
The thermal shift $\delta G \equiv \langle G^2\rangle_T - \langle G^2\rangle_0$
is evaluated using the trace anomaly relation, which is renormalization-group
invariant and gives, at leading order in
$\alpha_s$~\cite{Collins:1976yq,Nielsen:1977sy,Hatsuda:1990uw},
\begin{align}
  \delta G
  =
  -(4\pi^2)\frac{8}{9}
  \left[
  \epsilon-3p
  -
  \sum_{f=u,d,s}m_f\,
  \delta\langle\bar{f}f\rangle_T
  \right]\ .
  \label{eq:ope_gluon_condensate_trace_anomaly}
\end{align}
We decompose this into the trace-anomaly and quark-condensate contributions,
$\delta G = \delta G^{(I)} + \delta G^{(q)}$, where
\begin{align}
\delta G^{(I)}
&=
-(4\pi^2)\frac{8}{9}\,I\ ,
\qquad
I \equiv \epsilon-3p\ ,
\\
\delta G^{(q)}
&=
(4\pi^2)\frac{8}{9}
\sum_{f=u,d,s}m_f\,
\delta\langle\bar{f}f\rangle_T\ .
\end{align}
For the trace anomaly $I$, we use the smooth parameteraizion fitted to the lattice QCD equation of state in Refs.~\cite{Borsanyi:2010cj,Borsanyi:2013bia},
\begin{align}
  \frac{I(T)}{T^4}
  =
  \exp\!\left(
    -\frac{h_1}{t}
    -\frac{h_2}{t^2}
  \right)
  \left[
    h_0
    +
    \frac{
      f_0\left\{\tanh(f_1 t+f_2)+1\right\}
    }{
      1+g_1 t+g_2 t^2
    }
  \right]\ ,
\end{align}
where $t=T/200~\mathrm{MeV}$, with the parameter set of 
Ref.~\cite{Borsanyi:2013bia},
\begin{align}
  h_0 &= 0.1396\ ,
  &
  h_1 &= -0.1800\ ,
  &
  h_2 &= 0.0350\ ,
  \nonumber\\
  f_0 &= 2.76\ ,
  &
  f_1 &= 6.79\ ,
  &
  f_2 &= -5.29\ ,
  \nonumber\\
  g_1 &= -0.47\ ,
  &
  g_2 &= 1.04\ .
\end{align}
The quark-condensate contribution $\delta G^{(q)}$ is evaluated using the
lattice-based parametrizations of Eqs.~\eqref{Eq:para_1}
and~\eqref{Eq:para_2}, with the quark masses listed in
table~\ref{tab:ope_input_parameters}~\cite{PDG2024}.

\subsubsection{Four-Quark Scalar Condensates: $\langle j^2\rangle,\ \langle j_5^2\rangle,\ \langle \bar{s}js\rangle$}

The dimension-six four-quark condensates are evaluated using the
vacuum-saturation approximation ansatz~\cite{Kim:2019ybi,Shifman:1978bx,Shifman:1978by}, in which a
four-quark condensate is approximated as a product of two-quark
condensates. This approximation is motivated by large-$N_c$ factorization
in vacuum. 
We apply this approximation also at finite temperature and obtain
\begin{align}
  \langle j^2\rangle_T
  &= 0\ ,
  \label{eq:ope_j2_factorization}
  \\
  \langle j_5^2\rangle_T
  &=
  \frac{16}{36}\,(4\pi\alpha_s)\,
  \langle\bar{s}s\rangle_T^2,
  \label{eq:ope_j5_factorization}
  \\
  \langle\bar{s}js\rangle_T
  &=
  -\frac{16}{36}\,(4\pi\alpha_s)\,
  \langle\bar{s}s\rangle_T^2\ .
  \label{eq:ope_sjs_factorization}
\end{align}
The first condensate is neglected following Ref.~\cite{Kim:2019ybi},
as it is of higher order in $\alpha_s$ than the others.

\subsection{Non-Scalar Condensates}
\label{subsec:input_nonscalar_condensates}

This subsection specifies the numerical inputs for the non-scalar 
condensates entering the OPE at dimensions four and six.
At finite temperature, the four-velocity $u^\mu$ allows spin-two and 
spin-four operators to acquire non-vanishing expectation values, which 
vanish in vacuum by Lorentz symmetry.
We classify the condensates into three groups: the dimension-four spin-two 
condensates $(F^{\mu\nu}, G_2^{\mu\nu})$, the dimension-six spin-two 
condensates $(X_{\alpha\beta}, Y_{\alpha\beta}, Z_{\alpha\beta}, 
H_{\alpha\beta})$, and the dimension-six spin-four condensates 
$(K_{\alpha\beta\gamma\delta}, G_{4\alpha\beta\gamma\delta}, 
A_{\alpha\beta\gamma\delta}, B_{\alpha\beta\gamma\delta}, 
C_{\alpha\beta\gamma\delta})$.
All of these are estimated within the dilute pion-gas approximation.
Below the QCD crossover temperature, the thermal medium at vanishing baryon density is dominated by pions, and the dilute pion-gas approximation provides a reasonable description.

\subsubsection{Dilute Pion-Gas Approximation}

At low temperature, the thermal medium is approximated by a dilute gas 
of pions. Throughout, we use the charged-pion mass listed in 
table~\ref{tab:ope_input_parameters} as a common mass.
For a generic operator $\mathcal{O}$, the thermal expectation value is
\begin{align}
  \langle \mathcal{O}\rangle_T
  &=
  \langle 0|\mathcal{O}|0\rangle
  \nonumber\\
  &+
  \sum_{A=1}^{d_\pi}
  \int
  \frac{d^3k}{2E_{\pi,k}(2\pi)^3}\,
  \langle \pi^A(\mathbf{k})|\mathcal{O}|\pi^A(\mathbf{k})\rangle\,
  n_B(E_{\pi,k}/T)\ ,
  \label{eq:ope_meson_gas_condensate_master_reorg}
\end{align}
where $d_\pi=3$ and $E_{\pi,k}=\sqrt{\mathbf{k}^2+m_\pi^2}$.
The one-pion matrix elements are parametrized in terms of the pion momentum $p^\mu$ as
\begin{align}
  \langle \pi^A(p)|\mathcal{O}|\pi^A(p)\rangle
  &=
  C_{\mathcal{O},0}^{\pi}\ ,
\\
  \langle \pi^A(p)|\mathcal{O}_{\mu\nu}|\pi^A(p)\rangle
  &=
  C_{\mathcal{O},2}^{\pi}\,
  \mathcal{ST}(p_\mu p_\nu)\ ,
\\
  \langle \pi^A(p)|\mathcal{O}_{\mu_1\mu_2\mu_3\mu_4}|\pi^A(p)\rangle
  &=
  C_{\mathcal{O},4}^{\pi}\,
  \mathcal{ST}(p_{\mu_1}p_{\mu_2}p_{\mu_3}p_{\mu_4})\ .
\end{align}
Substituting them into Eq.~\eqref{eq:ope_meson_gas_condensate_master_reorg} 
and performing the momentum integral gives
\begin{align}
\delta\langle \mathcal{O}\rangle_T
  &=
  d_\pi\, C_{\mathcal{O},0}^{\pi}\,
  K_0(T;m_\pi)\ ,
\\
\langle \mathcal{O}_{\mu\nu}\rangle_T
  &=
  d_\pi\, C_{\mathcal{O},2}^{\pi}\,
  K_2(T;m_\pi)\,
  \mathcal{ST}(u_\mu u_\nu)\ ,
\label{Eq:cond_index2}\\
\langle \mathcal{O}_{\mu_1\mu_2\mu_3\mu_4}\rangle_T
  &=
  d_\pi\, C_{\mathcal{O},4}^{\pi}\,
  K_4(T;m_\pi)\,
  \mathcal{ST}(u_{\mu_1}u_{\mu_2}u_{\mu_3}u_{\mu_4})\ ,\label{Eq:cond_index4}
\end{align}
with
\begin{align}
  K_0(T;m)
  &=
  \frac{T^2}{24}\,
  B_1\!\left(\frac{m}{T}\right)\ ,
\\
  K_2(T;m)
  &=
  \frac{1}{360}
  \left[
    8\pi^2T^4B_2\!\left(\frac{m}{T}\right)
    -
    5m^2T^2B_1\!\left(\frac{m}{T}\right)
  \right]\ ,
\\
  K_4(T;m)
  &=
  \frac{32\pi^4}{315}\,T^6\,B_3\!\left(\frac{m}{T}\right)
  \nonumber\\
  &-\frac{\pi^2}{25}\,m^2T^4\,B_2\!\left(\frac{m}{T}\right)
  +\frac{1}{120}\,m^4T^2\,B_1\!\left(\frac{m}{T}\right)\ ,
\\
B_n(x)
&=
\frac{1}{\zeta(2n)\,\Gamma(2n)}
\int_x^\infty dy\,
y^{2(n-1)}
\frac{\sqrt{y^2-x^2}}{e^y-1}\ .
\end{align}
In Eqs.~\eqref{Eq:cond_index2} and~\eqref{Eq:cond_index4}, the non-scalar condensates vanish in vacuum by Lorentz invariance, 
and thus no $\delta$ is needed for them.
The thermal expectation value of each condensate is then determined by 
specifying the coefficient $C^{\pi}_{\mathcal{O},n}$ for the corresponding operator.

\subsubsection{Dimension-Four Spin-Two Condensates}
\label{subsubsec:input_dimension_four_spin_two_condensates}

The dimension-four spin-two condensates are
\begin{align}
F^{\mu\nu}
&=
\left\langle\mathcal{ST}\,
\bar{s}\gamma^\mu iD^\nu s
\right\rangle_T
\nonumber\\
&=
d_\pi\,
C^\pi_{F,2}
\,K_2(T;m_\pi)\,
\mathcal{ST}\!\left(u^\mu u^\nu\right)\ ,
\label{eq:ope_twist2_strange_definition_reorg}
\\
G_2^{\mu\nu}
&=
\left\langle
g^2\,
\mathcal{ST}\,
G^{a\mu}_{\ \ \alpha}G^{a\nu\alpha}
\right\rangle_T
\nonumber\\
&=
d_\pi\,
C^\pi_{G,2}
\,K_2(T;m_\pi)\,
\mathcal{ST}\!\left(u^\mu u^\nu\right)\ .
\label{eq:ope_twist2_gluon_definition_reorg}
\end{align}
The coefficients are taken from Ref.~\cite{Gubler:2018ctz}, 
\begin{align}
C^\pi_{F,2}
&=
A_2^{\pi(s)}|_{\mu=1\text{GeV}}
=
0.0257\ ,
\\
C^\pi_{G,2}
&=
2(4\pi \alpha_s)\times A_2^{\pi(g)}|_{\mu=1\text{GeV}}
=
2(4\pi \alpha_s)\times0.380\ ,
\end{align}
where they are derived from moments of the pion parton distribution 
functions of Ref.~\cite{Gluck:1999xe}.

\subsubsection{Dimension-Six Spin-Two Condensates}
\label{subsubsec:input_dimension_six_spin_two_condensates}

The dimension-six spin-two condensates involve gluon field strengths with
two covariant derivatives,
\begin{align}
X_{\alpha\beta}
  &=
\left\langle
g^2\,
  \mathcal{ST}\,
  G^a_{\mu\nu}D_\beta D_\alpha G^{a\mu\nu}
\right\rangle
\nonumber\\
&=
  d_\pi\,
C_{X,2}^{\pi}\, K_2(T;m_\pi)\, \mathcal{ST}(u_\alpha u_\beta)\ ,
  \label{eq:ope_X_definition}
\\
  Y_{\alpha\beta}
  &=
\left\langle
g^2\,
  \mathcal{ST}\,
  G^a_{\alpha\mu}D^\mu D^\nu G^a_{\beta\nu}
\right\rangle
\nonumber\\
&=
  d_\pi\,
 C_{Y,2}^{\pi}\, K_2(T;m_\pi)\, \mathcal{ST}(u_\alpha u_\beta)\ ,
  \label{eq:ope_Y_definition}
\\
  Z_{\alpha\beta}
  &=
\left\langle
g^2\,
  \mathcal{ST}\,
  G^a_{\alpha\mu}D_\beta D_\nu G^{a\mu\nu}
\right\rangle
\nonumber\\
&=
  d_\pi\,
 C_{Z,2}^{\pi}\, K_2(T;m_\pi)\, \mathcal{ST}(u_\alpha u_\beta)\ .
  \label{eq:ope_Z_definition}
\end{align}
To determine the coefficients, we evaluate the one-pion matrix elements
using the prescription proposed in Ref.~\cite{Lee:1997zta}: each covariant
derivative acting on a gluon field is replaced by the average gluon four-momentum inside the pion,
\begin{align}
  D_\mu G_{\rho\sigma}
  \simeq
  -i R^{(\pi)}_g\, p_\mu\, G_{\rho\sigma}\ ,
  \label{eq:ope_gluon_derivative_ansatz}
\end{align}
where $R^{(\pi)}_g$ is the gluon momentum fraction in the pion, taken
from the lattice-QCD determination at $\mu = 2$~GeV~\cite{ExtendedTwistedMass:2024kjf},
\begin{align}
  R^{(\pi)}_g = 0.388(49)\ .
\end{align}
Since no determination at $\mu=1~\mathrm{GeV}$ is currently available,
this value is used as a reference input also for the $\mu=1~\mathrm{GeV}$
analysis.
Applying the replacement twice gives
\begin{align}
  D_\beta D_\alpha G_{\rho\sigma}
  \simeq
  -(R^{(\pi)}_g)^2\, p_\beta p_\alpha\, G_{\rho\sigma}\ ,
  \label{eq:ope_two_derivative_ansatz}
\end{align}
and analogous formulas hold for $D^\mu D^\nu G_{\beta\nu}$ and
$D_\beta D_\nu G^{\mu\nu}$. Under this approximation, the coefficients
are
\begin{align}
C^\pi_{X,2}
&=
-(R^{(\pi)}_g)^2\, C^\pi_{G,0}\ ,\\
C^\pi_{Y,2}
&=
(R^{(\pi)}_g)^2\, \frac{C^\pi_{G,0}+3C^\pi_{G,2}m_\pi^2}{12}\ ,
\\
C^\pi_{Z,2}
&=
3C^\pi_{Y,2}\ ,
\end{align}
where $C^\pi_{G,0}$ is the one-pion matrix element of the gluon scalar
condensate~\cite{Hatsuda:1992bv},
\begin{align}
C^\pi_{G,0} = -\frac{32\pi^2 m_\pi^2}{9}\ .
\end{align}

The remaining dimension-six spin-two condensate $H_{\alpha\beta}$ 
vanishes under the vacuum-saturation approximation. Under this 
approximation, the tensor before the symmetric traceless projection 
takes the form
\begin{align}
\left\langle
g^2 \left(\bar{q} t^a \gamma_5 \gamma_\alpha q\right)
\left(\bar{q} t^a \gamma_5 \gamma_\beta q\right)
\right\rangle
\propto
g_{\alpha\beta}\,\langle\bar{q}q\rangle^2\ ,
\end{align}
which is removed by the $\mathcal{ST}$ projection, giving 
$H_{\alpha\beta} = 0$.

\subsubsection{Dimension-Six Spin-Four Condensates}
\label{subsubsec:input_dimension_six_spin_four_condensates}

Among the dimension-six spin-four condensates, the condensates
\(K_{\alpha\beta\gamma\delta}\) and \(G_{4\alpha\beta\gamma\delta}\),
given by
\begin{align}
K_{\alpha\beta\gamma\delta}
&=
\left\langle
\mathcal{ST}\,
\bar{q} \gamma_\alpha D_\beta D_\gamma D_\delta q
\right\rangle_T
\nonumber\\
&=
  d_\pi\,
C^\pi_{K,4}\,
K_4(T;m_\pi)\,
\mathcal{ST}
\!\left(u^\alpha u^\beta u^\gamma u^\delta \right)\ ,
\\
G_{4\alpha\beta\gamma\delta}
&=
\left\langle
g^2\,
\mathcal{ST}\,
G^a_{\alpha\mu} D_\delta D_\gamma G^{a}_{\beta}{}^\mu
\right\rangle_T
\nonumber\\
&=
  d_\pi\,
C^\pi_{G,4}\,
K_4(T;m_\pi)\,
\mathcal{ST}
\!\left(u^\alpha u^\beta u^\gamma u^\delta \right)\ ,
\end{align}
are evaluated using the same procedure as that used for the dimension-four spin-two condensates.  
This procedure is based on moments
of the pion parton distribution functions of Ref.~\cite{Gluck:1999xe}.
The corresponding coefficients are given in
Ref.~\cite{Gubler:2018ctz} as
\begin{align}
C^\pi_{K,4}
&=
iA_4^{\pi(s)}|_{\mu=1\text{GeV}}
=i\,0.00154\ ,\\
C^\pi_{G,4}
&=
-2(4\pi \alpha_s)\times  A_4^{\pi(g)}|_{\mu=1\text{GeV}}
=
-2(4\pi \alpha_s)\times 0.0593\ .
\end{align}

The one-pion matrix elements of the remaining spin-four condensates 
$A_{\alpha\beta\gamma\delta}$, $B_{\alpha\beta\gamma\delta}$, and 
$C_{\alpha\beta\gamma\delta}$ are not known, and we neglect these 
contributions,
\begin{align}
A_{\alpha\beta\gamma\delta}
=
B_{\alpha\beta\gamma\delta}
=
C_{\alpha\beta\gamma\delta}
=
0.
\end{align}
This omission introduces a uncertainty in the analysis. 
However, within the OPE framework, dimension-six operator contributions 
are expected to be numerically small, and their absence is unlikely to 
qualitatively affect the results.

\subsection{Phenomenological Modeling of the Spectral Function}
\label{subsec:ch4_hadronic_spectral_ansatz}

This subsection specifies the phenomenological ansatz for the spectral 
function appearing on the right-hand side of 
Eq.~\eqref{eq:reduced_dispersion_relation}.
For each polarization channel $X=L,T$, the spectral function is modeled as a sum of two distinct contributions,
\begin{align}
  \rho_{\text{X}}(s,v^2;T)
  =
  \rho_{\text{X},\mathrm{res}}(s,v^2;T)
  +
  \rho_{\text{X},\mathrm{high}}(s,v^2;T),
  \label{eq:tilde_rho_X_ansatz_res_cont}
\end{align}
where the first term represents the $\phi$ meson resonance, while the second term represents the contribution from higher-energy states.
In general, thermal contributions such as Landau damping could also 
enter the spectral function~\cite{Hatsuda:1992bv}. In the present case, a relevant contribution may arise from the $K\bar{K}$ loop, which lies outside the 
scope of the pion-gas approximation adopted here; we therefore neglect 
it. We also note that thermal contributions of this type can carry 
support at $s=0$, which would introduce an additional term in the 
dispersion relation~\cite{Klingl:1997kf,Leupold:1998bt}.

The $\phi$meson resonance is modeled by a delta function,
\begin{align}
  \rho_{\text{X},\mathrm{res}}^{\phi}(s,v^2;T)
  =
  f_{\phi,\text{X}}^2(T,v^2)\,
  \delta\!\left(s-m_{\phi,\text{X}}^2(T,v^2)\right),
  \label{eq:tilde_rho_phi_res_delta}
\end{align}
where $m_{\phi,\text{X}}$ and $f_{\phi,\text{X}}$ denote the polarization-dependent 
mass and decay constant, respectively. 
This ansatz 
is motivated by the narrow vacuum width of the $\phi$ meson 
($\Gamma_\phi \simeq 4~\mathrm{MeV}$~\cite{PDG2024}); 
model analyses suggest that the temperature-induced broadening remains 
at most $\sim 40~\mathrm{MeV}$ at $T = 150~\mathrm{MeV}$ and vanishing 
baryon density~\cite{Lissauer:1991fr,Ko:1993id,Haglin:1994ap,Vujanovic:2009wr,Kumar:2020vys,Kaur:2025kjk}.

The contribution from higher-energy states is modeled as
\begin{align}
  \rho_{\text{X},\mathrm{high}}(s,v^2;T)
  =
  \frac{1}{\pi}\mathrm{Im}\,\Pi^{\mathrm{scalar}}_{\mathrm{pert}}(s)\,
  \theta(s-s_{0,\text{X}})\ ,
\end{align}
where $s_{0,\text{X}}$ is the threshold marking the onset of continuous
higher-energy states. 
The use of $\frac{1}{\pi}\mathrm{Im}\,\Pi^{\mathrm{scalar}}_{\mathrm{pert}}$ 
is motivated by the reliability of the perturbative description at large 
$s$, and is supported empirically by its successful application in QCD 
sum rule analyses~\cite{Gubler:2018ctz}.
This perturbative correlator $\Pi^{\mathrm{scalar}}_{\mathrm{pert}}(s)$ is obtained by the analytic continuation from the
spacelike expression,
\begin{align}
\Pi^{\mathrm{scalar}}_{\mathrm{pert}}(Q^2)
&=
-\frac{1+\alpha_s/\pi}{4\pi^2}\ln\frac{Q^2}{\mu^2}
-\frac{3m_s^2}{2\pi^2 Q^2}
\nonumber\\
&-\frac{m_s^2}{\pi^2}\,\frac{\alpha_s}{\pi}\,
\frac{4-3\ln(Q^2/\mu^2)}{Q^2}\ ,\quad (Q^2>0)
\label{eq:had_perturbative_scalar_input}
\end{align}
via $Q^2 = -s - i0$.
The resulting continuum spectral function is
\begin{align}
\rho_{\text{X},\mathrm{high}}(s,v^2;T)
=
\left[
\frac{1+\alpha_s(\mu)/\pi}{4\pi^2}
+
\frac{3m_s^2\alpha_s(\mu)}{\pi^3 s}
\right]\theta(s-s_{0,\text{X}})\ .
\end{align}

\subsection{Borel Transformaion}
\label{subsec:borel_sum_rule}

The Borel transformation is applied to both sides of the dispersion 
relation in Eq.~\eqref{eq:reduced_dispersion_relation} to efficiently extract 
information on the $\phi$ meson resonance while suppressing the contributions from higher-energy 
states.
The Borel transform of a function $f(Q^2)$ is defined by
\begin{align}
  \mathcal{B}_{M^2}\left[f\right]
  &\equiv
  \lim_{\substack{Q^2,N\to\infty\\ Q^2/N=M^2}}
  \frac{(Q^2)^N}{(N-1)!}
  \left(-\frac{d}{dQ^2}\right)^N f(Q^2)\ .
  \label{eq:borel_operator_definition}
\end{align}
This transformation serves two purposes. First, it eliminates polynomial 
terms in $Q^2$; as a result, the subtraction terms and the polynomial 
part of the surface contribution are removed.
Second, $1/(s-q^2)$ 
in the dispersion integral is replaced by $(1/M^2)\,e^{-s/M^2}$, which 
exponentially suppresses the high-energy states contribution and enhances the 
sensitivity to the low-lying $\phi$ meson
resonance~\cite{Shifman:1998rb}.

Explicitly, applying the Borel transformation to 
Eq.~\eqref{eq:reduced_dispersion_relation} gives
\begin{align}
\mathcal{B}_{M^2}\!\left[\Pi_{\text{X}}(q^2,v^2)\right]
=
\Pi^\text{res}_{\text{X}}(M^2,v^2)
+
\Pi^\text{high}_{\text{X}}(M^2,v^2)\ .
\label{eq:borel_dispersion_relation}
\end{align}
where the resonance term is 
\begin{align}
\Pi^\text{res}_{\text{X}}(M^2,v^2)&=  \frac{1}{M^2}
  \int_{-v^2}^{\infty} ds\,
  \rho_{\text{X},\mathrm{res}}\,e^{-s/M^2}
\nonumber\\
&=
  \frac{f_{\phi,\text{X}}^2(T,v^2)}{M^2}
  \exp\!\left[-\frac{m_{\phi,\text{X}}^2(T,v^2)}{M^2}\right]\ ,
\end{align}
while high-energy states part is
\begin{align}
\Pi^\text{high}_{\text{X}}(M^2,v^2)
&=  \frac{1}{M^2}
  \int_{-v^2}^{\infty} ds\,
  \rho_{\text{X},\mathrm{cont}}\,e^{-s/M^2}
\nonumber\\
&=
  \frac{1+\alpha_s(\mu)/\pi}{4\pi^2}\,
  e^{-s_{0,\text{X}}(T,v)/M^2}
\nonumber\\
&  +
  \frac{3m_s^2\alpha_s(\mu)}{\pi^3 M^2}\,
  E_1\!\left(\frac{s_{0,\text{X}}(T,v)}{M^2}\right)\ ,
\end{align}
with
\begin{align}
  E_1(x)
  \equiv
  \int_x^\infty dt\,\frac{e^{-t}}{t}
  =
  \Gamma(0,x)\ .
\end{align}
When $s_{0,\text{X}}\gg m_{\phi,\text{X}}^2$, the two contributions are well separated, and the sum rule provides a reliable 
constraint on parameters in the vector-meson spectral function. 
Explicit forms of the Borel-transformed OPE side of Eq.~\eqref{eq:borel_dispersion_relation}
is quite long, and is given in Appendix~\ref{app:borel_ope_expressions} (which is the same expression as shown in Ref.~\cite{Kim:2019ybi}).

\section{Numerical results}\label{sec:results}

We present the numerical results of the $\phi$ meson properties 
at finite temperature.
We first describe the zero-momentum results, where the longitudinal and transverse 
channels are degenerate.
We then study the finite-momentum dependence and analyze the longitudinal-transverse 
splitting for different temperatures. 
Finally, we examine the origin of the splitting by selectively removing thermal 
contributions in the operator product expansion.

\subsection{Numerical analysis method}
\label{subsec:numerical_analysis_method}

Before presenting the numerical results, we summarize the procedure used to
extract the $\phi$ meson mass and decay constant from the Borel sum rule.
The analysis is based on the rearranged form of Eq.~(\ref{eq:borel_dispersion_relation}),
\begin{align}
  A_{\text{X}}(M^2,v^2;s_0)
  =
  \Pi^\text{res}_{\text{X}}(M^2,v^2)\ ,
\end{align}
with
\begin{align}
  A_{\text{X}}(M^2,v^2;s_0)
  \equiv
  \mathcal{B}_{M^2}\!\left[\Pi_{\text{X}}(q^2,v^2)\right]
  -\Pi^\text{high}_{\text{X}}(M^2,v^2)\ .
\end{align}
For each temperature $T$, momentum $v$, and channel $X = L, T$,
we independently determine the optimal continuum threshold
$s_{0,\text{X}}^{\star}(T,v)$ and extract the corresponding $\phi$ meson mass and decay constant.

As the first step, we determine the so-called Borel window 
\begin{align}
  M_{\min,\text{X}}(v^2,T)
  \leq M \leq
  M_{\max,\text{X}}(v^2,T;s_0)\ ,
\end{align}
for given $s_0$, $T$, and $v$. 
The window, within which the analysis is considered reliable, is defined by two conditions.
The lower boundary $M_{\min,\text{X}}$ is set by the OPE convergence condition,
\begin{align}
  C_{\text{X}}(M^2,v^2)
  =
  \left|
  \frac{
    \mathcal{B}_{M^2}\!\left[\Pi^{(d=6)}_{\text{X}}(q^2,v^2)\right]
  }{
    \mathcal{B}_{M^2}\!\left[\Pi_{\text{X}}(q^2,v^2)\right]
  }
  \right|
  \leq 0.1\ ,
\end{align}
where $\Pi^{(d=6)}_{\text{X}}$ denotes the dimension-six contribution to
$\Pi_{\text{X}}(q^2,v^2)$.
The upper boundary $M_{\max,\text{X}}$ is set by the pole-dominance condition,
\begin{align}
  P_{\text{X}}(M^2,v^2;s_0)
  =
  \left|
  \frac{
    A_{\text{X}}(M^2,v^2;s_0)
  }{
    \mathcal{B}_{M^2}\!\left[\Pi_{\text{X}}(q^2,v^2)\right]
  }
  \right|
  \geq 0.5\ .
\end{align}
If a Borel window satisfying both the OPE-convergence and pole-dominance
conditions cannot be defined, the corresponding trial value of \(s_0\)
is discarded.

Within the Borel window, 
the $\phi$ meson mass and its decay constant for generic $M^2$ are extracted as
\begin{align}
  m_{\phi,\text{X}}^2(M^2,v^2;s_0)
  &=
  M^2
  -
  \frac{
    dA_{\text{X}}(M^2,v^2;s_0)/d(M^{-2})
  }{
    A_{\text{X}}(M^2,v^2;s_0)
  }\ ,\\
  f_{\phi,\text{X}}^2(M^2,v^2;s_{0,\text{X}})
  &=
  M^2\,
  A_{\text{X}}(M^2,v^2;s_{0,\text{X}})
\nonumber\\
&\quad\times
  \exp\!\left[
    \frac{m^2_{\phi,\text{X}}(M^2,v^2;s_{0,\text{X}})}{M^2}
  \right] .
\end{align}
The Borel-window averages are then
\begin{align}
  \bar{m}_{\phi,\text{X}}(v^2;s_0)
  &=
  \frac{
    \int_{M_{\min,\text{X}}}^{M_{\max,\text{X}}}
    dM\, m_{\phi,\text{X}}(M^2,v^2;s_0)
  }{
    M_{\max,\text{X}} - M_{\min,\text{X}}
  }\ ,\\
  \bar{f}_{\phi,\text{X}}(v^2;s_0)
  &=
  \frac{
    \int_{M_{\min,\text{X}}}^{M_{\max,\text{X}}}
    dM\, f_{\phi,\text{X}}(M^2,v^2;s_0)
  }{
    M_{\max,\text{X}} - M_{\min,\text{X}}
  }\ ,
\end{align}
where the arguments of $M_{\min,\text{X}}$ and $M_{\max,\text{X}}$ are omitted for notational simplicity.
Stability of the extracted mass with respect to the Borel mass is estimated with a variance
\begin{align}
  &\chi_{m,\text{X}}^2(v^2;s_0)
  \nonumber\\
  &\quad=
  \frac{
    \int_{M_{\min,\text{X}}}^{M_{\max,\text{X}}}
    dM\,
    \left[
      m_{\phi,\text{X}}(M^2,v^2;s_0)
      -
      \bar{m}_{\phi,\text{X}}(v^2;s_0)
    \right]^2
  }{
    M_{\max,\text{X}} - M_{\min,\text{X}}
  }\ .
\end{align}
This quantity serves as a stability measure of the Borel-mass dependence
and is not a statistical chi-square.
The continuum threshold is determined as the value of $s_0$ that minimizes $\chi_{m,\text{X}}^2$, i.e., 
\begin{align}
  s_{0,\text{X}}^{\star}(T,v^2)
  =
  \arg\min_{s_0}\,
  \chi_{m,\text{X}}^2(T,v^2;s_0)\ .
\end{align}
The $\phi$meson mass and its decay constant are then determined as
\begin{align}
  m_{\phi,\text{X}}(T,v^2)
  &=
  \bar{m}_{\phi,\text{X}}
  \!\left(T,v^2;\,s_{0,\text{X}}^{\star}(T,v^2)\right)\ ,\\
  f_{\phi,\text{X}}(T,v^2)
  &=
  \bar{f}_{\phi,\text{X}}
  \!\left(T,v^2;\,s_{0,\text{X}}^{\star}(T,v^2)\right)\ .
\end{align}

\subsection{Temperature dependence at zero momentum}
\label{subsec:results_Tdep_vzero}

\begin{figure}[htbp]
  \centering
  \includegraphics[width=0.48\textwidth]{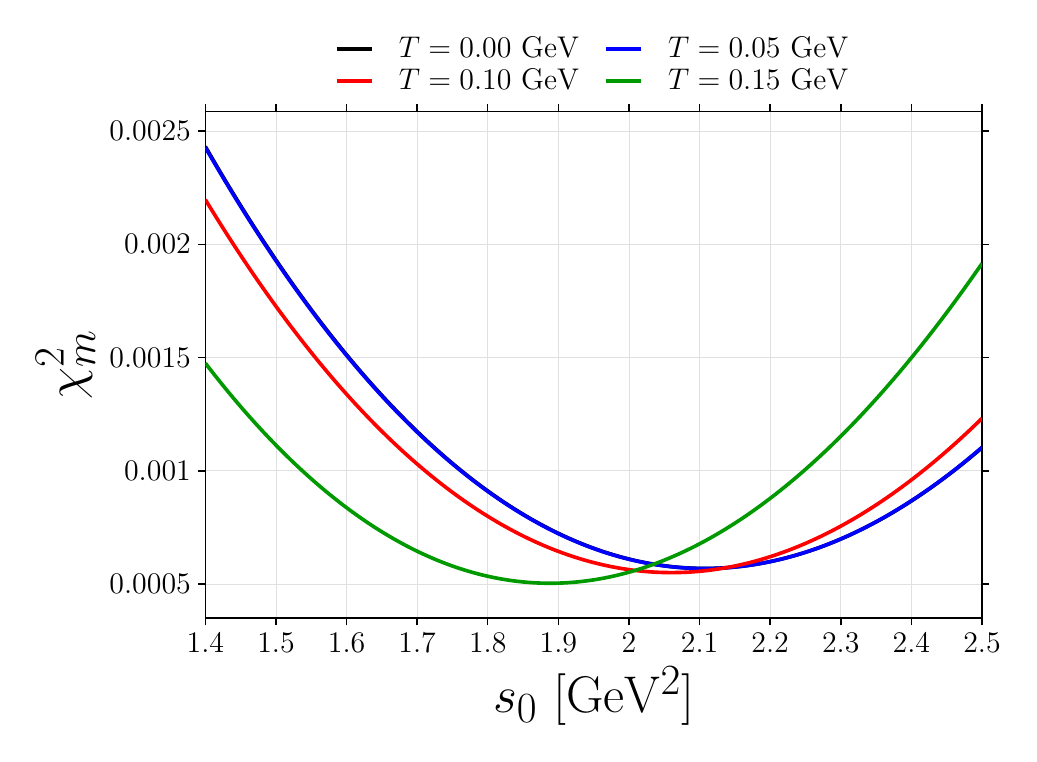}
  \caption{
Stability measure $\chi^2_m$ as a function of the continuum threshold $s_0$
at zero momentum for representative temperatures $T = 0.0,\, 0.05,\, 0.1,\, 0.15~\mathrm{GeV}$.
The minimum of each curve determines the optimal continuum threshold $s_0^\star$.  }
  \label{fig:chi_s0_vzero}
\end{figure}

We first consider the zero-momentum case, $v=|\mathbf{q}|=0$.
In this limit, the longitudinal and transverse channels are degenerate, and the analysis reduces to that for a single scalar amplitude.
Figure~\ref{fig:chi_s0_vzero} shows the stability measure $\chi^2_m=\chi^2_{m,\text{X}}(v^2=0;s_0)$ as a function of the continuum threshold $s_0$ at representative temperatures $T=0.0,\,0.05,\,0.1,\,0.15$ GeV.
The minimum of $\chi^2_m$ determines the optimal threshold $s_0^\star$ used in the subsequent analysis. In practice, the scan over $s_0$ is performed with a grid spacing of $\Delta s_{0,\text{X}}=0.001$ GeV$^2$, which is sufficiently small compared with the overall scale of variation of $\chi^2_m$. In the figure, $\chi^2_m$ shows little variation in the range $T=0.0$--$0.05$ GeV, while for $T=0.05$--$0.15$ GeV the position of the minimum is seen to shift progressively toward smaller values of $s_0$.

\begin{figure}[htbp]
  \centering
  \includegraphics[width=0.48\textwidth]{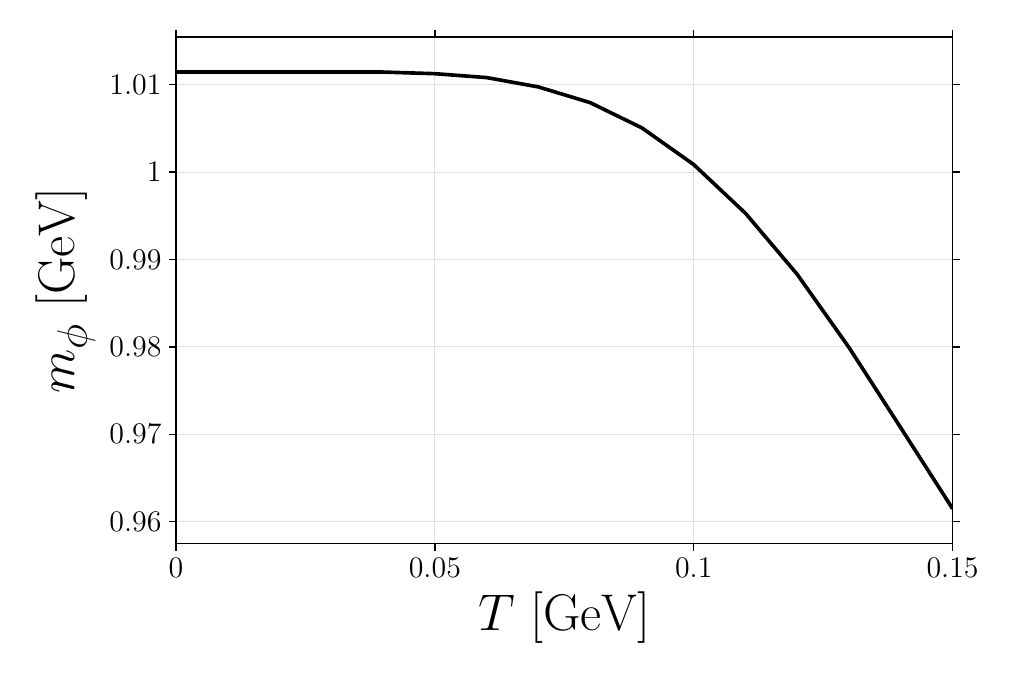}
  \includegraphics[width=0.48\textwidth]{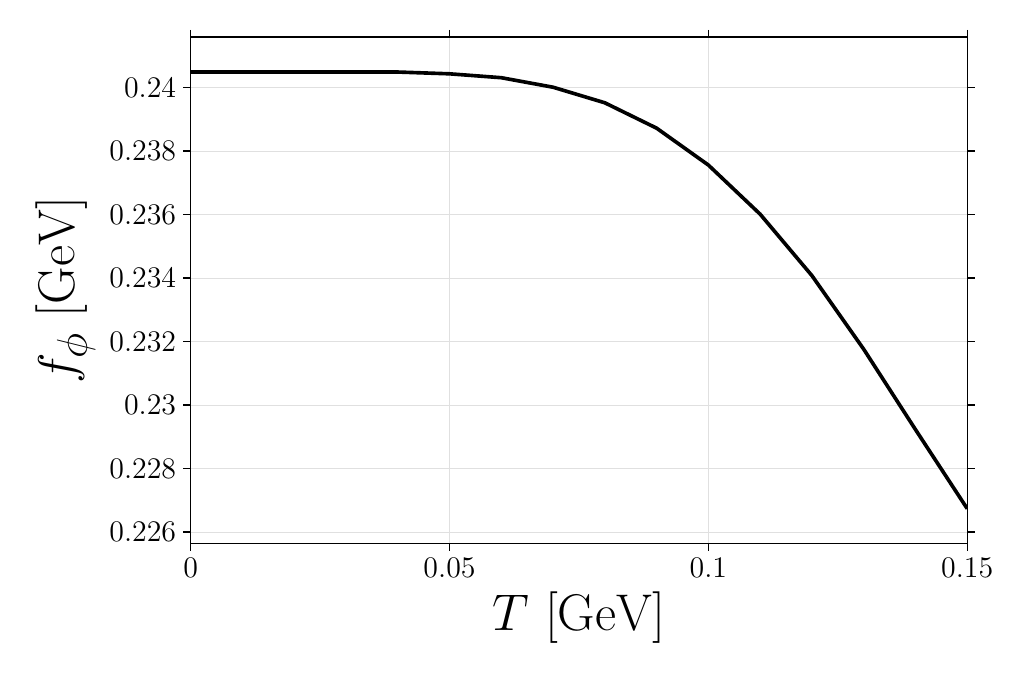}
  \includegraphics[width=0.48\textwidth]{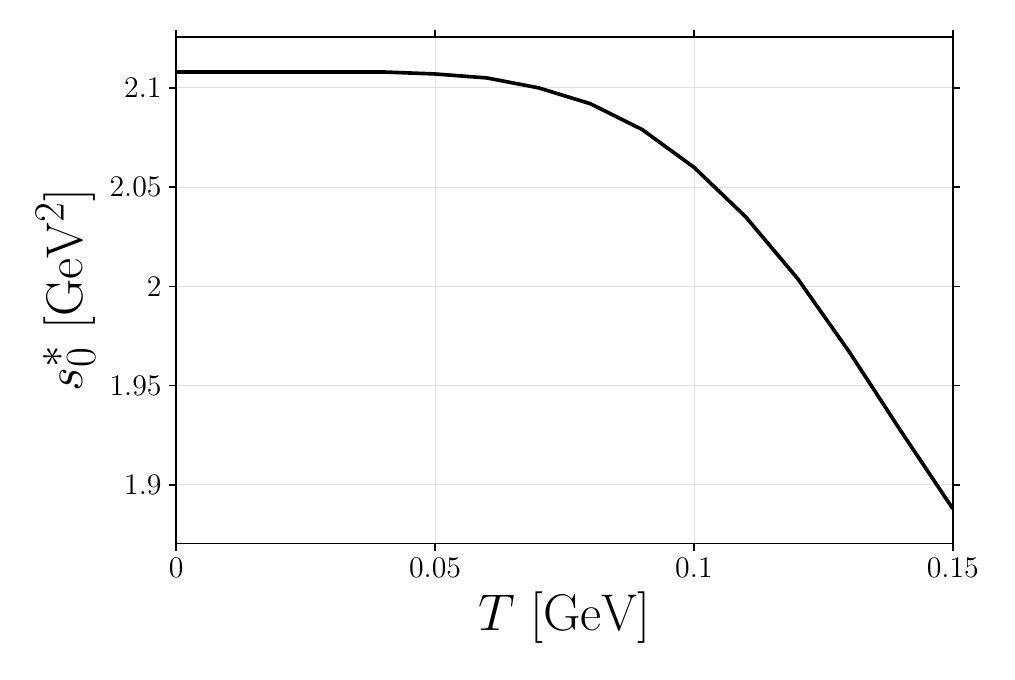}
  \caption{
  Temperature dependence of the pole parameters at zero momentum.
  The panels show the $\phi$meson mass $m_\phi$, the residue $f_\phi$, and the
  continuum threshold $s_0$.
  The temperature range is
  $0 \leq T \leq 0.15~\mathrm{GeV}$.
  }
  \label{fig:Tdep_vzero}
\end{figure}

Using the optimized value of $s_0$, we next study the temperature dependence at zero momentum. 
Figure~\ref{fig:Tdep_vzero} shows the temperature dependence of $m_\phi$, $f_\phi$, and $s^\star_0$ in the range $0~\mathrm{GeV} \leq T \leq 0.15~\mathrm{GeV}$.
All three quantities remain nearly constant for $T \lesssim 0.06~\mathrm{GeV}$ and then decrease at higher temperatures.
All the three quantities exhibits a monotonic dependence on temperature for $T \gtrsim 0.06~\mathrm{GeV}$. 
This correlated behavior was also reported in the QCD sum rule analysis
of the $\phi$ meson in nuclear matter~\cite{Kim:2019ybi}

\subsection{Momentum dependence and transverse and longitudinal splitting}
\label{subsec:results_vdep_TL}

\begin{figure}[htbp]
  \centering
  \includegraphics[width=0.48\textwidth]{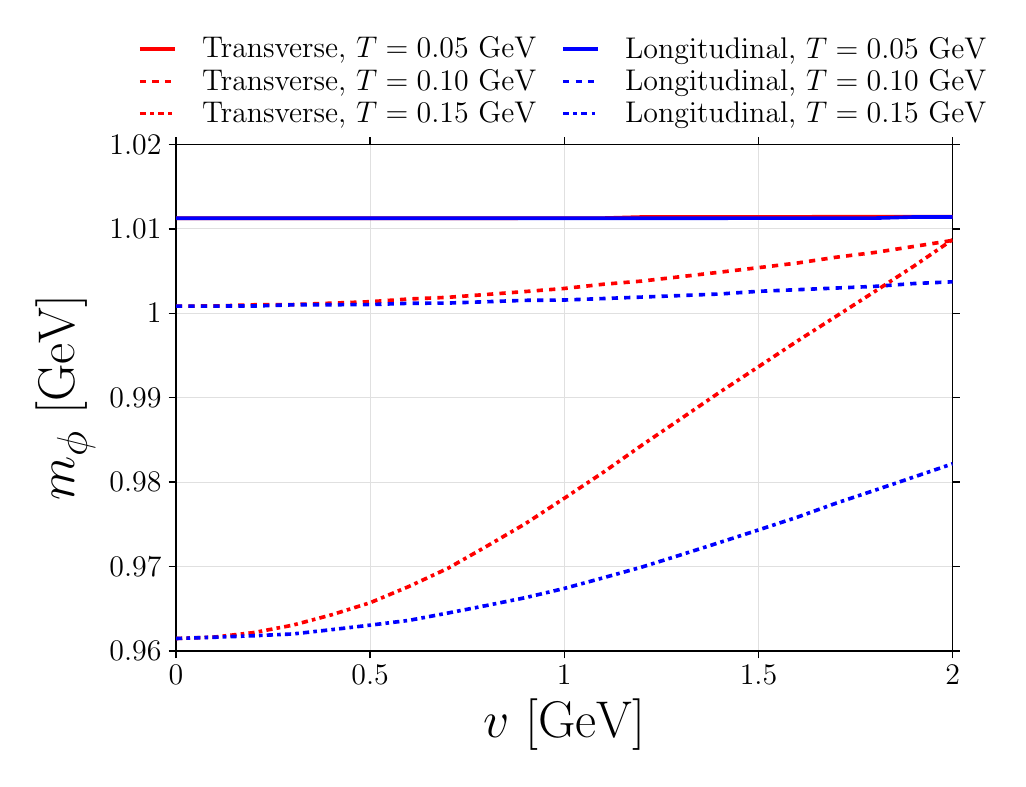}  %\includegraphics[width=0.48\textwidth]{figure7_splitting_mphi_T010_continuous_borel_readable.pdf}
  \caption{
Momentum dependence of the $\phi$meson mass at fixed temperatures.
Red and blue lines show the transverse and longitudinal channels, respectively.
Solid, dashed, and dash-dotted lines correspond to
$T=0.05$, $0.10$, and $0.15~\mathrm{GeV}$, respectively.
  }
  \label{fig:mass_vdep}
\end{figure}

We next turn to the finite-momentum regime $v \neq 0$, where the longitudinal 
and transverse channels of the current-current correlator, defined with respect 
to the medium rest frame, become distinct. Consequently, the properties of the $\phi$ meson should be determined separately for the two channels.
Figure~\ref{fig:mass_vdep} shows the momentum dependence of the $\phi$-meson
mass for the transverse and longitudinal channels at $T = 0.05$, $0.10$, and
$0.15~\mathrm{GeV}$.
At small momentum, the two channels are nearly degenerate, confirming the
expected degeneracy in the low-momentum limit.
As $v$ increases, both masses rise and a clear splitting develops between them,
with the transverse mass exceeding the longitudinal mass.
The magnitude of this splitting grows with temperature. 

The finite-temperature mass splitting obtained here for $T = 0.1$ GeV is much smaller than the
nuclear matter counterpart reported in the QCD sum rule analysis
of Ref.~\cite{Kim:2019ybi} and the recent quark-meson coupling model analysis of Ref.~\cite{Arifi:2026eus}, 
where the longitudinal-transverse splitting reaches the order of tens of MeV. 
However, at $T = 0.15$ GeV, the splitting is of the same order of magnitude 
as that obtained in the above works for dense matter at normal nuclear matter density. 
This tendency is consistent with theoretical calculations based on hadronic
models that systematically estimate both nuclear matter and finite-temperature
effects within a unified framework~\cite{Kaur:2025kjk,Vujanovic:2009wr,Kumar:2020vys}.

\subsection{Origin of the transverse and longitudinal splitting}
\label{subsec:results_origin_TL_splitting}

\begin{figure}[htbp]
  \centering
  \includegraphics[width=0.48\textwidth]{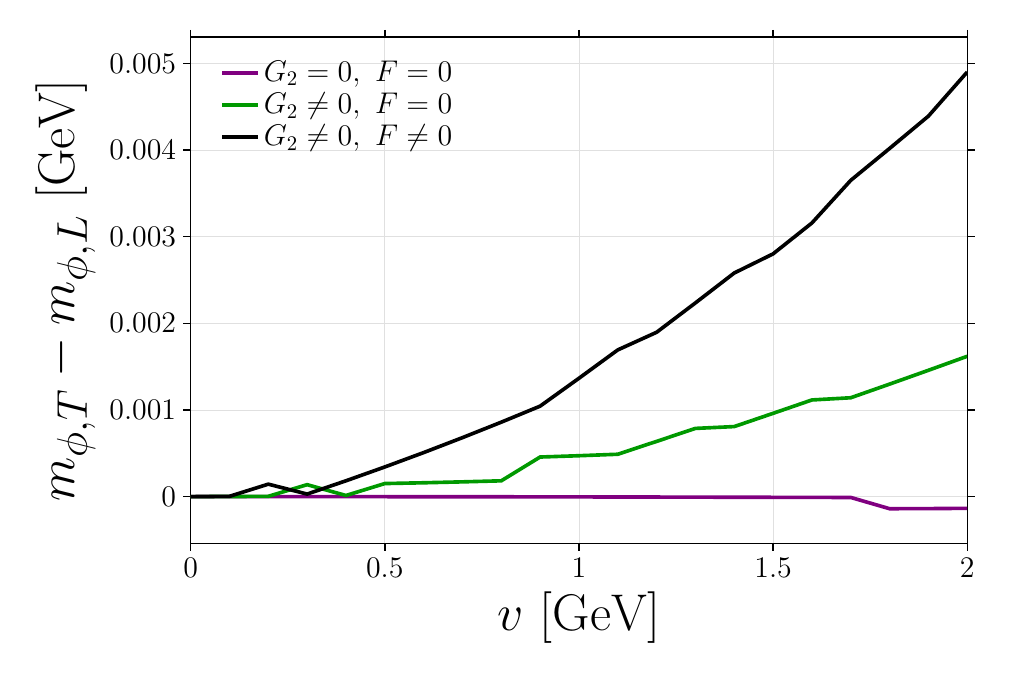}
  \caption{Comparison of the transverse--longitudinal mass splitting of the $\phi$ meson 
with and without the dimension-four condensate terms $F$ and $G_2$ at 
$T=0.10~\mathrm{GeV}$. The black, green, and purple lines show the results 
obtained by including both condensates, switching off the $F$ 
condensate alone, and switching off both the $F$ and $G_2$ condensates, 
respectively.}
  \label{fig:contribution_analysis_amp}
\end{figure}

\begin{figure}[htbp]
  \centering
  \includegraphics[width=0.48\textwidth]{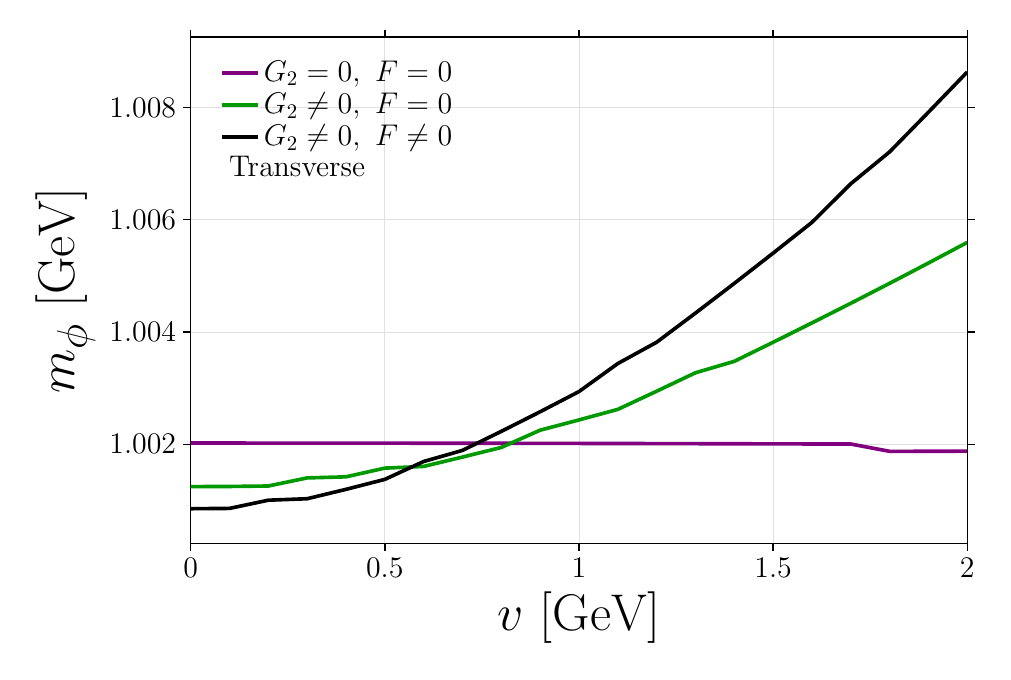}
  \includegraphics[width=0.48\textwidth]{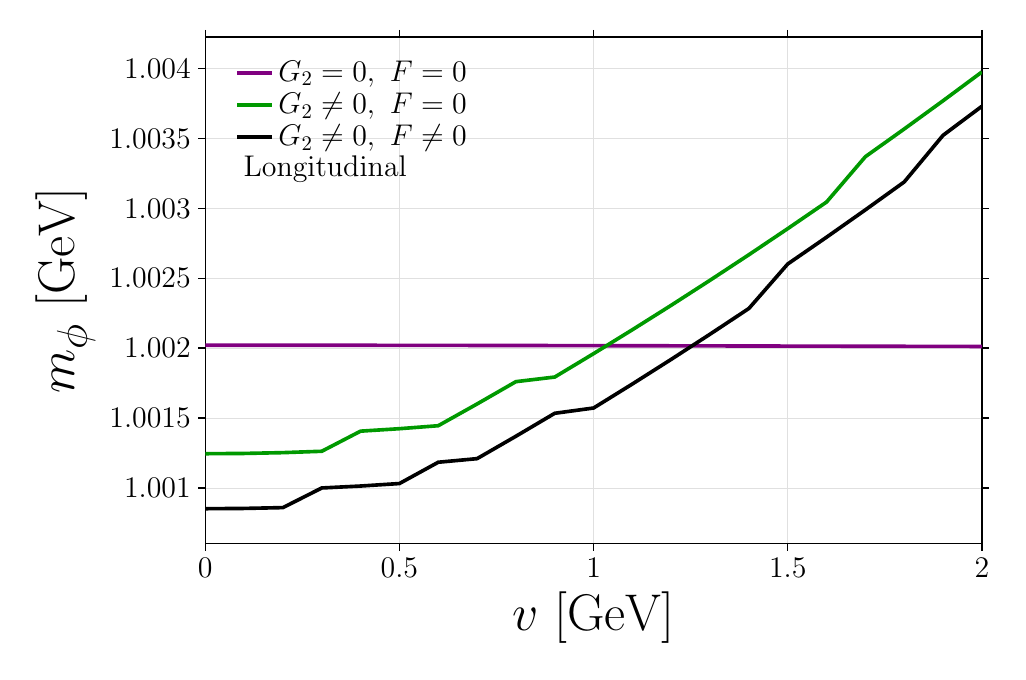}
  \caption{
Same as Fig.~\ref{fig:contribution_analysis_amp}, but the upper and lower panels show the transverse and longitudinal 
channels separately at $T=0.10~\mathrm{GeV}$, respectively.}
  \label{fig:contribution_analysis}
\end{figure}

Finally, we investigate which thermal contributions are responsible for the
transverse and longitudinal splitting.  To this end, we compare the full result with those from modified analyses in which selected leading tensor condensate terms in the OPE are
switched off. 
The leading tensor terms appear with the dimension-four quark condensate $F$ with one derivative and gluon condensate $G_2$, evaluated in Sec.~\ref{subsubsec:input_dimension_four_spin_two_condensates}. 
Here, we suppress the Lorentz indices on those condensates for notational simplicity. 

Figure~\ref{fig:contribution_analysis_amp} compares the transverse--longitudinal
mass splitting, $m_{\phi,T}-m_{\phi,L}$, at $T=0.10~\mathrm{GeV}$, among the
full result, the result obtained by switching off the $F$ condensate
alone, and the result obtained by switching off both the $F$ and
$G_2$ condensates. The full result shows a clear momentum-dependent splitting.
When the $F$ condensate is switched off, the splitting is clearly
reduced, and it is further suppressed when both the $F$ and $G_2$
condensates are switched off. This comparison indicates that the
transverse--longitudinal mass splitting is mainly driven by the dimension-four
condensates.

Figure~\ref{fig:contribution_analysis} shows the momentum dependence of the $\phi$ meson mass at $T=0.10~\mathrm{GeV}$ 
in the transverse and longitudinal channels. 
In each plot, we compare the full result with those 
obtained by selectively switching off 
the dimension-four thermal condensates, $F$ and $G_2$. 
In the transverse channel, as shown in
the upper panel, both condensates 
contribute comparably in magnitude to the upward mass shift at large momentum.
In the longitudinal channel, as shown in the lower panel,  switching off the $F$ condensate has a relatively minor effect on the result. 
The residual major momentum dependence stems from the $G_2$ condensate. 
These comparisons indicate that the $G_2$ condensate gives the dominant contribution to the mass shift in the longitudinal channel, while the $F$ condensate also gives a comparable contribution in the
transverse channel.

\section{Summary}\label{sec:summary}
We have studied the $\phi$ meson at finite temperature and finite momentum using
QCD sum rules.
The strange vector current correlator has been decomposed into the transverse and
longitudinal channels with respect to 
the medium flow vector.
The operator product expansion has been evaluated up to dimension six, including
scalar condensates and non-scalar condensates.
The non-scalar condensates, 
which have purely thermal origin,
have been estimated
in the dilute pion-gas approximation.
At zero momentum, the transverse and longitudinal channels are degenerate.
In this limit, we have found that the optimized continuum threshold, the $\phi$ meson
mass, and the decay constant remain nearly constant for $T \lesssim 0.06~\mathrm{GeV}$
and decrease for $0.06~\mathrm{GeV} \lesssim T \lesssim 0.15~\mathrm{GeV}$.

At finite momentum, the transverse and longitudinal channels split.
We have found that both masses increase with momentum, with the transverse mass
exceeding the longitudinal mass. 
Also, we have found that the size of this splitting grows with temperature. 
We have further investigated the origin of the finite-momentum splitting by
selectively removing dimension-four thermal condensate contributions.
The comparison has shown that the $G_2$ condensate gives the dominant contribution to the mass shifts, especially in the longitudinal channel,
while both the $G_2$ and $F$ condensates give visible contributions
to the transverse--longitudinal difference.

As a future direction, it is important to connect the finite-temperature
modifications of the transverse and longitudinal spectral functions obtained
in this work to experimentally accessible signals.
This requires a more explicit treatment of the dilepton and \(K\bar{K}\) decay
channels, which would allow one to estimate how the finite-temperature
transverse--longitudinal splitting appears in measured spectra. 
It is also interesting to examine effects from thermal kaon
excitations, including possible Landau-damping contributions.  
Although this effect is naively suppressed in the low-temperature range considered here due to a suppressed thermal distribution, it could be sizable because of a direct coupling to the \(\phi\) meson through strange hadronic channels.

\begin{acknowledgments}
This work is supported by the National Natural Science Foundation of China (NSFC) under grant numbers W2433010 and W2532002 and 
by JSPS KAKENHI under grant numbers JP25H00400, JP26K07113, and 23K22487.
\end{acknowledgments}

\appendix

\appendix

\section{Borel-transformed OPE expressions}
\label{app:borel_ope_expressions}

In this appendix, we summarize the explicit Borel-transformed OPE
expressions used in the numerical analysis.  
For convenience, we define
\begin{align}
  \mathcal I_n(M^2)
  &\equiv
  \mathcal B_{M^2}
  \left[
    \frac{1}{(Q^2)^n}
  \right]
  =
  \frac{1}{(n-1)!}\,
  \frac{1}{(M^2)^n}\ ,
  \label{eq:app_borel_In_definition}
  \\
  \mathcal L_n(M^2)
  &\equiv
  \mathcal B_{M^2}
  \left[
    \frac{L_Q}{(Q^2)^n}
  \right]
  =
  \left[
    \ln\frac{M^2}{\mu^2}
    +
    \psi(n)
  \right]\mathcal I_n(M^2)\ ,
  \label{eq:app_borel_Ln_definition}
  \\
  \mathcal K_n(M^2)
  &\equiv
  \mathcal B_{M^2}
  \left[
    \frac{L_Q^2}{(Q^2)^n}
  \right]
  \nonumber\\
  &=
  \left[
    \left(
      \ln\frac{M^2}{\mu^2}
      +
      \psi(n)
    \right)^2
    -
    \dot{\psi}(n)
  \right]\mathcal I_n(M^2)\ ,
  \label{eq:app_borel_Kn_definition}
\end{align}
where $L_Q=\ln(Q^2/\mu^2)$ and $\psi(n)$ is the digamma function.

The longitudinal OPE after the Borel transformation is
\begin{widetext}
\begin{align}
  \mathcal B_{M^2}\!\left[\Pi_{\text{L}}(q^2,v^2)\right]
  &=
  \frac{1+\alpha_s/\pi}{4\pi^2}
  -\frac{3m_s^2}{2\pi^2}\mathcal I_1
  -\frac{m_s^2}{\pi^2}\frac{\alpha_s}{\pi}
  \left(4\mathcal I_1-3\mathcal L_1\right)
  \nonumber\\
  &\quad
  +2m_s\langle\bar s s\rangle \mathcal I_2
  +\frac{2m_s\alpha_s}{3\pi}\langle \bar s s\rangle \mathcal I_2
  -\frac{8m_s^3\langle\bar s s\rangle}{3}\mathcal I_3
  -\frac{4\langle\bar s j s\rangle}{9}\mathcal I_3
  -2\langle j_5^2\rangle\mathcal I_3
  \nonumber\\
  &\quad
  +\frac{\langle G^2\rangle}{\pi^2}
  \left(
    \frac{1}{48}\mathcal I_2
    +\frac{m_s^2}{36}\mathcal I_3
  \right)
  +\frac{7\alpha_s}{288\pi^4}\langle G^2\rangle \mathcal I_2
  +\frac{3m_s^4}{4\pi^2}\left(2\mathcal L_2-\mathcal I_2\right)
  \nonumber\\
  &\quad
  -\frac{m_s^4}{6\pi^2}\frac{\alpha_s}{\pi}
  \Bigl(
    32\mathcal I_2
    -24\zeta(3)\mathcal I_2
    -33\mathcal L_2
    +18\mathcal K_2
  \Bigr)
  \nonumber\\
  &\quad
  +\left(
    2\mathcal I_2
    -3m_s^2\mathcal I_3
    -6m_s^2v^2\mathcal I_4
  \right)\widehat F
  \nonumber\\
  &\quad
  +\frac{\widehat G_2}{48\pi^2}
  \Bigl[
    -4\mathcal L_2
    -16m_s^2\mathcal L_3
    +48m_s^2 v^2\mathcal L_4
    +9\mathcal I_2
    -11m_s^2\mathcal I_3
    -8v^2\mathcal I_3
    -22m_s^2 v^2\mathcal I_4
  \Bigr]
  \nonumber\\
  &\quad
  +\frac{\widehat X}{480\pi^2}
  \Bigl[
    25\mathcal L_3
    -30v^2\mathcal L_4
    +20\mathcal I_3
    -32v^2\mathcal I_4
  \Bigr]
  +\frac{\widehat Y}{2880\pi^2}
  \Bigl[
    -90\mathcal L_3
    +300v^2\mathcal L_4
    -245\mathcal I_3
    +86v^2\mathcal I_4
  \Bigr]
  \nonumber\\
  &\quad
  +\frac{\widehat Z}{960\pi^2}
  \Bigl[
    150\mathcal L_3
    -180v^2\mathcal L_4
    +165\mathcal I_3
    -182v^2\mathcal I_4
  \Bigr]
  +\frac{2i}{3}
  \left(
    5\mathcal I_3
    -6v^2\mathcal I_4
  \right)\widehat K
  \nonumber\\
  &\quad
  +\frac{\widehat G_4}{4320\pi^2}
  \Bigl[
    -330\mathcal L_3
    +396v^2\mathcal L_4
    +1025\mathcal I_3
    -2238v^2\mathcal I_4
    +1152v^4\mathcal I_5
  \Bigr].
  \label{eq:app_borel_PiL_full}
\end{align}
\end{widetext}

The transverse OPE after the Borel transformation is
\begin{widetext}
\begin{align}
  \mathcal B_{M^2}\!\left[\Pi_{\text{T}}(q^2,v^2)\right]
  &=
  \frac{1+\alpha_s/\pi}{4\pi^2}
  -\frac{3m_s^2}{2\pi^2}\mathcal I_1
  -\frac{m_s^2}{\pi^2}\frac{\alpha_s}{\pi}
  \left(4\mathcal I_1-3\mathcal L_1\right)
  \nonumber\\
  &\quad
  +2m_s\langle\bar s s\rangle \mathcal I_2
  +\frac{2m_s\alpha_s}{3\pi}\langle \bar s s\rangle \mathcal I_2
  -\frac{8m_s^3\langle\bar s s\rangle}{3}\mathcal I_3
  -\frac{4\langle\bar s j s\rangle}{9}\mathcal I_3
  -2\langle j_5^2\rangle\mathcal I_3
  \nonumber\\
  &\quad
  +\frac{\langle G^2\rangle}{\pi^2}
  \left(
    \frac{1}{48}\mathcal I_2
    +\frac{m_s^2}{36}\mathcal I_3
  \right)
  +\frac{7\alpha_s}{288\pi^4}\langle G^2\rangle \mathcal I_2
  +\frac{3m_s^4}{4\pi^2}\left(2\mathcal L_2-\mathcal I_2\right)
  \nonumber\\
  &\quad
  -\frac{m_s^4}{6\pi^2}\frac{\alpha_s}{\pi}
  \Bigl(
    32\mathcal I_2
    -24\zeta(3)\mathcal I_2
    -33\mathcal L_2
    +18\mathcal K_2
  \Bigr)
  \nonumber\\
  &\quad
  +\left(
    2\mathcal I_2
    -4v^2\mathcal I_3
    -3m_s^2\mathcal I_3
    +9m_s^2 v^2\mathcal I_4
  \right)\widehat F
  \nonumber\\
  &\quad
  +\frac{\widehat G_2}{48\pi^2}
  \Bigl[
    -4\mathcal L_2
    -16m_s^2\mathcal L_3
    +8v^2\mathcal L_3
    +8m_s^2 v^2\mathcal L_4
    +9\mathcal I_2
    -11m_s^2\mathcal I_3
    -14v^2\mathcal I_3
    +33m_s^2 v^2\mathcal I_4
  \Bigr]
  \nonumber\\
  &\quad
  +\frac{\widehat X}{480\pi^2}
  \Bigl[
    25\mathcal L_3
    -35v^2\mathcal L_4
    +20\mathcal I_3
    -24v^2\mathcal I_4
  \Bigr]
  +\frac{\widehat Y}{2880\pi^2}
  \Bigl[
    -90\mathcal L_3
    +30v^2\mathcal L_4
    -245\mathcal I_3
    +447v^2\mathcal I_4
  \Bigr]
  \nonumber\\
  &\quad
  +\frac{\widehat Z}{960\pi^2}
  \Bigl[
    150\mathcal L_3
    -210v^2\mathcal L_4
    +165\mathcal I_3
    -239v^2\mathcal I_4
  \Bigr]
  +\frac{2i}{3}
  \left(
    5\mathcal I_3
    -27v^2\mathcal I_4
    +24v^4\mathcal I_5
  \right)\widehat K
  \nonumber\\
  &\quad
  +\frac{\widehat G_4}{4320\pi^2}
  \Bigl[
    -330\mathcal L_3
    +1782v^2\mathcal L_4
    -1584v^4\mathcal L_5
    +1025\mathcal I_3
    -5031v^2\mathcal I_4
    +4344v^4\mathcal I_5
  \Bigr].
  \label{eq:app_borel_PiT_full}
\end{align}
\end{widetext}

For reference, we write down 
the explicit form of $\mathcal I_n$ ($n=1,2,\cdots,5$),
%\begin{align}
%  \mathcal I_1(M^2) &= \frac{1}{M^2}\ ,\quad
%  \mathcal I_2(M^2) = \frac{1}{(M^2)^2}\ ,\\
%  \mathcal I_3(M^2) &= \frac{1}{2(M^2)^3}\ ,\quad
%  \mathcal I_4(M^2) = \frac{1}{6(M^2)^4}\ ,\\
%  \mathcal I_5(M^2) &= \frac{1}{24(M^2)^5}\ .
%  \label{eq:app_borel_In_explicit}
%\end{align}
\begin{align}
  \mathcal I_1(M^2) &= \frac{1}{M^2}\ ,\\
  \mathcal I_2(M^2) &= \frac{1}{(M^2)^2}\ ,\\
  \mathcal I_3(M^2) &= \frac{1}{2(M^2)^3}\ ,\\
  \mathcal I_4(M^2) &= \frac{1}{6(M^2)^4}\ ,\\
  \mathcal I_5(M^2) &= \frac{1}{24(M^2)^5}\ .
  \label{eq:app_borel_In_explicit}
\end{align}
The logarithmic functions $\mathcal L_n$ needed above are
\begin{align}
  \mathcal L_1(M^2)
  &=
  \left(
    \ln\frac{M^2}{\mu^2}
    -\gamma_E
  \right)
  \frac{1}{M^2}\ ,
  \\
  \mathcal L_2(M^2)
  &=
  \left(
    \ln\frac{M^2}{\mu^2}
    +1-\gamma_E
  \right)
  \frac{1}{(M^2)^2}\ ,
  \\
  \mathcal L_3(M^2)
  &=
  \left(
    \ln\frac{M^2}{\mu^2}
    +\frac{3}{2}-\gamma_E
  \right)
  \frac{1}{2(M^2)^3}\ ,
  \\
  \mathcal L_4(M^2)
  &=
  \left(
    \ln\frac{M^2}{\mu^2}
    +\frac{11}{6}-\gamma_E
  \right)
  \frac{1}{6(M^2)^4}\ ,
  \\
  \mathcal L_5(M^2)
  &=
  \left(
    \ln\frac{M^2}{\mu^2}
    +\frac{25}{12}-\gamma_E
  \right)
  \frac{1}{24(M^2)^5}\ .
  \label{eq:app_borel_Ln_explicit}
\end{align}
We also use
\begin{align}
  \mathcal K_2(M^2)
  =
  \left[
    \left(
      \ln\frac{M^2}{\mu^2}
      +1-\gamma_E
    \right)^2
    -
    \left(
      \frac{\pi^2}{6}-1
    \right)
  \right]
  \frac{1}{(M^2)^2}\ .
  \label{eq:app_borel_K2_explicit}
\end{align}

\bibliography{apssamp}% Produces the bibliography via BibTeX.

\end{document}